\documentstyle[preprint,aps,tighten,prd,epsf]{revtex}
\newcommand{\IoI}{I\otimes I}
\newcommand{\gdoxd}[2]{\gamma_{#1}^\dagger\otimes\xi_{#2}^\dagger}
\newcommand{\ggox}[3]{\gamma_{#1}\gamma_{#2}\otimes\xi_{#3}}
\newcommand{\ggoI}[2]{\gamma_{#1}\gamma_{#2}\otimes I}
\newcommand{\ggoxx}[4]{\gamma_{#1}\gamma_{#2}\otimes\xi_{#3}\xi_{#4}}
\newcommand{\goI}[1]{\gamma_{#1}\otimes I}
\newcommand{\gox}[2]{\gamma_{#1}\otimes\xi_{#2}}
\newcommand{\goxx}[3]{\gamma_{#1}\otimes\xi_{#2}\xi_{#3}}
\begin{document}
\draft
\preprint{UTHEP-413}
\title{Pion decay constant for the Kogut-Susskind quark action in
       quenched lattice QCD}
\author{S. Aoki$^1$,
        M. Fukugita$^2$,
        S. Hashimoto$^3$,
        K-I. Ishikawa$^3$,
        N. Ishizuka$^{1, 4}$,
        Y. Iwasaki$^{1, 4}$,
        K. Kanaya$^{1, 4}$,
        T. Kaneda$^1$,
        S. Kaya$^3$,
        Y. Kuramashi$^{3, 5}$,
        M. Okawa$^3$,
        T. Onogi$^6$,
        S. Tominaga$^3$,
        N. Tsutsui$^6$,
        A. Ukawa$^{1, 4}$,
        N. Yamada$^6$,
        and
        T. Yoshi\'e$^{1, 4}$}
\address{(JLQCD Collaboration) \\
         $^1$Institute of Physics, University of Tsukuba,
         Tsukuba, Ibaraki 305-8571, Japan \\
         $^2$Institute for Cosmic Ray Research, University
         of Tokyo, Tanashi, Tokyo 188-8502,Japan \\
         $^3$High Energy Accelerator Research Organization
         (KEK), Tsukuba, Ibaraki 305-0801, Japan \\
         $^4$Center for Computational Physics, University
         of Tsukuba, Tsukuba, Ibaraki 305-8577, Japan \\
         $^5$Department of Physics, Washington University,
         St. Louis, Missouri 63130, USA \\
         $^6$Department of Physics, Hiroshima University,
         Higashi-Hiroshima, Hiroshima 739-8526, Japan}
\date{\today}
\maketitle
\begin{abstract}
We present a study for the pion decay constant $f_\pi$ in the quenched
approximation to lattice QCD with the Kogut-Susskind (KS) quark action,
with the emphasis given to the renormalization problems.  Numerical
simulations are carried out at the couplings $\beta = 6.0$ and 6.2 on
$32^3\times 64$ and $48^3\times 64$ lattices, respectively.  The pion
decay constant is evaluated for all KS flavors via gauge invariant and
non-invariant axial vector currents with the renormalization constants
calculated by both non-perturbative method and perturbation theory.  We
obtain $f_\pi = 89(6)$ MeV in the continuum limit as the best value
using the partially conserved axial vector current, which requires no
renormalization.  From a study for the other KS flavors we find that the
results obtained with the non-perturbative renormalization constants are
well convergent among the KS flavors in the continuum limit, confirming
restoration of $\rm SU(4)_A$ flavor symmetry, while perturbative
renormalization still leaves an apparent flavor breaking effect even in
the continuum limit.
\end{abstract}

\widetext

\section{Introduction}

In recent large-scale simulations of lattice QCD, statistical errors of
physical quantities have become quite small.  Indeed, for some hadronic
matrix elements the precision has been so high that we cannot ignore
uncertainties coming from the renormalization factor of lattice
operators.  Thus it has become increasingly important to reduce
uncertainties from this source.

Renormalization factors can be evaluated in perturbation theory.
Pushing the calculation beyond the one-loop level is usually difficult,
however, hence uncertainties arising from higher-order corrections
remain.  We expect this problem of higher-order uncertainties to profit
fully from a non-perturbative treatment.  A non-perturbative method for
calculating renormalization factors was proposed\cite{Mart95}, and has
been applied to the quark mass\cite{NPqmas}, decay constants\cite{NPbili} and
four-fermion operators\cite{NPfour}, with the Wilson and the clover quark
actions and to the quark mass with the Kogut-Susskind quark action%
\cite{JLQC99}.

An important point to check with non-perturbatively calculated
renormalization factors is their reliability and the degree of
improvement achieved in the final physical results.  For this purpose
the pion decay constant is perhaps the best choice because the reference
experimental value is known to a high precision.  A verification that
non-perturbative determination works for simple quark bilinear operators
is a first step to ensure validity of more general applications to four-%
quark or other operators.

In this work, the pion decay constant is examined with the Kogut-%
Susskind (KS) quark action via gauge invariant and non-invariant
operators using all KS flavors.  The KS action has the well-known
feature that $\rm SU(4)_A$ flavor symmetry is broken down to
$\rm U(1)_A$ subgroup at a finite lattice spacing.  We orient our
study mainly toward the following two points provided by this feature.
First, due to the remaining $\rm U(1)_A$ symmetry, the renormalization
constant for the corresponding axial vector current equals exactly unity,
and hence the pion decay constant calculated in this channel receives no
renormalization.  This makes it possible to attain a high-precision
calculation of the pion decay constant without uncertainties from
renormalization.  Second, we can calculate the pion decay constant using
axial vector currents in the other KS flavor channels.  Symmetry is
broken in the decay constants at a finite lattice spacing, but
restoration is expected in the continuum limit.  Such restoration of
full flavor symmetry has been previously examined for pion mass%
\cite{Shar92,JLQC97}.  Here we extend the study to the pion decay
constant, the new feature being the necessity of renormalization
constants.  This can be used to investigate the reliability of non-%
perturbative methods for the calculation of renormalization factors,
compared to perturbative treatments.  We also compare the results
obtained with gauge invariant operators to those with non-invariant ones.

The paper is organized as follows.  In Sec.~\ref{sec:formulations} we
establish our notations and formalism.  The method employed for our
calculations is explained in Sec.~\ref{sec:extraction}, followed by
discussion of perturbative and non-perturbative renormalization factors
in Sec.~\ref{sec:renormalization}.  We summarize the simulation details
in Sec.~\ref{sec:simulation}, and present the results on the chiral and
continuum extrapolations in Secs.~\ref{sec:chiral} and
\ref{sec:continuum}.  We close with a brief conclusion in Sec.~%
\ref{sec:conclusion}.
\pagebreak

\section{Formulations}
\label{sec:formulations}

\subsection{Kogut-Susskind quark action}

The Kogut-Susskind (KS) quark action is defined in terms of one-%
component fermion fields $\overline{\chi}(n)$ and $\chi(n)$ on a lattice
whose site is labelled by $n_\mu = 0, 1, 2,\ldots, L-1$,
\begin{eqnarray}
S_q^{\rm KS} = a^4\sum_n\left[\sum_\mu\eta_\mu(n)\frac{1}{2 a}\big(
\overline{\chi}(n)U_\mu(n)\chi(n +\hat{\mu})\right.\: \nonumber\\
-\overline{\chi}(n +\hat{\mu}) U^\dagger_\mu(n)\chi(n)\big) + m_q
\overline{\chi}(n)\chi(n)\left],\mbox{\rule{-10pt}{16pt}}\right.
\label{eq:action}
\end{eqnarray}
where $m_q$ is the bare quark mass and
$\eta_\mu = (-1)^{n_1+\cdots+n_{\mu-1}}$ is the KS sign factor.  Color
sums are assumed for simplicity.  Dividing the lattice into $2^4$-%
hypercubes which are labelled by $x_\mu = 0, 2, 4,\ldots, L-2$, and
whose corners are specified by a four-vector $A$ with $A_\mu = 0$ or 1,
we introduce sixteen-component fields
\begin{equation}
\overline{\phi}_A(x) =\frac{1}{4}\overline{\chi}(x + A), \qquad
\phi_A(x) =\frac{1}{4}\chi(x + A).
\end{equation}
In terms of these hypercubic fields, the action~(\ref{eq:action}) is
rewritten as
\begin{eqnarray}
S_q^{\rm KS} = (2 a)^4\sum_{x,AB}\left[\sum_\mu\big(
\overline{\phi}_A(x)(\goI{\mu})_{AB}\nabla_\mu U_{AB}\phi_B(x)
\mbox{\rule{0pt}{16pt}}\right.\;\;\;\;\, \nonumber\\
+ a\overline{\phi}_A(x)(\goxx{5}{\mu}{5})_{AB}\Delta_\mu U_{AB}
\phi_B(x)\big)\;\;\;\: \nonumber\\
\left.+ m_q\overline{\phi}_A(x)(\IoI)_{AB}U_{AB}(x, x)\phi_B(x)
\mbox{\rule{0pt}{16pt}}\right]. \nonumber\\
\label{eq:action2}
\end{eqnarray}
Here a hypercube matrix referring to the Dirac spinor
$\gamma_S =\gamma_1^{S_1}\cdots\gamma_4^{S_4}$ and the KS flavor
$\xi_F =\gamma_1^{*F_1}\cdots\gamma_4^{*F_4}$ is defined by
\begin{equation}
(\gox{S}{F})_{AB} =\frac{1}{4}{\rm Tr}(\gamma_A^\dagger\gamma_S
\gamma_B\gamma_F^\dagger)
\end{equation}
and the lattice derivatives are given by 
\begin{eqnarray}
\nabla_\mu U_{AB}\phi_B(x) & =
 &\:\frac{1}{2 a}\:\big(U_{AB}(x, y)\phi_B(y)|_{y=x+2\hat{\mu}}
  \nonumber\\
&&\;\;\:\:\:\, - U_{AB}(x, y)\phi_B(y) |_{y=x-2\hat{\mu}}\big), \\
\Delta_\mu U_{AB}\phi_B(x) & =
 &\frac{1}{2 a^2}\big(U_{AB}(x, y)\phi_B(y) |_{y=x+2\hat{\mu}}
  \nonumber\\
&&\;\;\;\;\; + U_{AB}(x, y)\phi_B(y) |_{y=x-2\hat{\mu}}
  \nonumber\\
&&\;\;\;\;\; - 2 U_{AB}(x, x)\phi_B(y)\big),
\end{eqnarray}
where $U_{AB}(x, y)$ is the average of ordered products of gauge link
variables over the shortest paths from $x + A$ to $y + B$.

A gauge-invariant meson operator with zero spatial momentum is defined
in this hypercubic notation by
\begin{equation}
O_S^F(x_4) =\sum_{\vec{x}}\sum_{AB}\overline{\phi}_A(x)(
\gox{S}{F})_{AB} U_{AB}(x, x)\phi_B(x).
\end{equation}
For instance, $O_S = A_\mu$, $\pi$, and $\rho_k$ respectively for
$\gamma_S =\gamma_\mu\gamma_5$, $\gamma_5$, and $\gamma_k$.  Here we
have $2^4\times 2^4 = 256$ operators, which are classified into
irreducible representations\cite{Golt86} in terms of
$d_\mu = S_\mu - F_\mu\,(\bmod{2})$.

The form of the action~(\ref{eq:action2}) shows that the flavor-mixing
term $(\goxx{5}{\mu}{5})$ breaks $\rm SU(4)_A$ flavor symmetry down to
$\rm U(1)_A$ subgroup for the flavor channel $\xi_5$ at a finite lattice
spacing.  A lattice analog of the PCAC relation holds in the $\xi_5$
channel corresponding to $\rm U(1)_A$ symmetry:
\begin{equation}
\nabla_\mu A_\mu^5(x) = 2 m_q\pi^5(x),
\label{eq:PCAC}
\end{equation}
where the superscript 5 refers to $\xi_5$.  On the other hand, there
appear additional terms in the PCAC relation for other channels, which
vanish only in the continuum limit.

\subsection{Pion decay constant}

The pion decay constant is defined in the continuum theory by
\begin{equation}
\sqrt{2} f_\pi m_\pi =\langle 0 |\overline{u}\gamma_4\gamma_5 d |
\pi^+(\vec{p} = 0)\rangle.
\end{equation}
We adopt the normalization $f_\pi^{\rm(exp)}\simeq 93$ MeV.  If we use
the PCAC relation, this may be rewritten as 
\begin{equation}
\sqrt{2} f_\pi m_\pi^2 = (m_u + m_d)\langle 0 |\overline{u}\gamma_5
d |\pi^+(\vec{p} = 0)\rangle.
\label{eq:PCAC2}
\end{equation}

The lattice pion decay constant for the KS flavor $\xi_F$ is defined by
\begin{equation}
\sqrt{2} f_\pi^F m_\pi^F =\langle 0 | A_4^F |\pi^F (\vec{p} = 0)
\rangle,
\label{eq:fpi}
\end{equation}
In the $\xi_5$ channel where the PCAC relation~(\ref{eq:PCAC}) holds, we
may use an alternative formula corresponding to Eq.~(\ref{eq:PCAC2}):
\begin{equation}
\sqrt{2} f_\pi^{(P)5} (m_\pi^5)^2 = 2 m_q\langle 0 |\pi^5 |\pi^5
(\vec{p} = 0)\rangle,
\label{eq:fpiP}
\end{equation}
where we have added the superscript $(P)$ to distinguish explicitly the
pion decay constant obtained with a pion operator from that with an
axial vector current.

\section{Extraction of pion decay constant}
\label{sec:extraction}

We employ the wall source technique to enhance signals\cite{Ishi94}.
The meson operator for the wall source at the origin is defined by
\begin{equation}
O_{SW}^F(0) =\sum_{\vec{x},\vec{y}}\overline{\phi}_A(\vec{x}, 0)(
\gox{S}{F})_{AB}\phi_B(\vec{y}, 0),
\end{equation}
where we assume that gauge configurations are fixed to some gauge.  The
matrix elements appearing in the definition of the pion decay constant
are extracted from the large-time behavior of the correlation function
at zero spatial momentum:
\begin{eqnarray}
\makebox[10pt][l]{$\langle O_S^F(t)\pi_W^F(0)\rangle$} &
 & \nonumber\\
&&\sim C_{O_S\pi_W}^F (\sigma_t)^t [\exp(- m_\pi^F t)\pm\exp(-
   m_\pi^F (T - t))], \nonumber\\
&&\hspace{95pt}\left\{
  \begin{array}{l}
   +\mbox{ sign for } O_S =\pi, \pi_W, \\
   -\mbox{ sign for } O_S = A_4,
  \end{array}
  \right.
  \label{eq:correlator}
\end{eqnarray}
where $m_\pi^F$ is pion mass common to the three cases.  Here we extend
the time slice of meson operator defined at
$x_4 = 0, 2, 4,\ldots, T - 2$ to have $t = 0, 1, 2,\ldots, T - 1$
extensions with the temporal lattice size $T$.  Note that there is no
mixing between $\sigma_t =\pm 1$ states in this case with time-extended
meson operators.  The amplitude $C_{O_S\pi_W}^F$ can be written up to an
overall sign factor as
\begin{equation}
C_{O_S\pi_W}^F =\frac{\langle 0 | O_S^F |\pi^F(\vec{p} = 0)\rangle
\langle\pi^F(\vec{p} = 0) |\pi_W^F | 0\rangle}{2 m_\pi^F V_s},
\end{equation}
with  $V_s$ the spatial lattice volume.  Using the amplitude of the
correlation functions with the axial vector current $(O_S = A_4)$ and
the pion operator for the wall source $(O_S = \pi_W)$, the pion decay
constant is calculated as
\begin{equation}
f_\pi^F =\frac{\sqrt{V_s}}{\sqrt{m_\pi^F}}
\frac{C_{A_4\pi_W}^F}{\sqrt{C_{\pi_W\pi_W}^F}},
\label{eq:fpi2}
\end{equation}
where the pion mass obtained by the correlation function with the pion
operator ($O_S = \pi$) is used in this work.  For comparison, the gauge
non-invariant axial vector current and pion operator to obtain the
amplitude and pion mass, respectively,
\begin{eqnarray}
{A_4'}\mbox{}^F(x_4) & =
 &\sum_{\vec{x}}\sum_{AB}\overline{\phi}_A(x)(
\ggox{4}{5}{F})_{AB}\phi_B(x), \\
{\pi'}\mbox{}^F(x_4) & =
 &\sum_{\vec{x}}\sum_{AB}\overline{\phi}_A(x)(
\gox{5}{F})_{AB}\phi_B(x),
\end{eqnarray}
is also examined.  Alternatively, an extraction of the decay constant
from the pion operator $(O_S^F =\pi^5)$ requires the combination given
by
\begin{equation}
f_\pi^{(P)5} =\frac{\sqrt{2m_q V_s}}{\sqrt{(m_\pi^5)^3}}
\frac{C_{\pi\pi_W}^5}{\sqrt{C_{\pi_W\pi_W}^5}}.
\label{eq:fpiP2}
\end{equation}

\section{Renormalization}
\label{sec:renormalization}

\subsection{General considerations}

Renormalization is necessary to extract the physical pion decay constant
from the lattice calculations.  This procedure is made for each flavor
in the case of the KS action.  It is expected that the renormalization
eliminates the KS flavor dependence in a way that the decay constant
calculated for various KS flavors takes a unique value in the continuum
limit.

Let us define a multiplicative renormalization constant $Z_A^F$ for the
lattice axial vector current $A_\mu^F\big|_{\rm lat}$ through
\begin{equation}
A_\mu^F\left|_{\rm phys} = Z_A^F A_\mu^F\right|_{\rm lat}.
\label{eq:ZA}
\end{equation}
According to the definition~(\ref{eq:fpi}) the pion decay constant
calculated with the axial vector current is renormalized as 
\begin{equation}
f_\pi^F\left|_{\rm phys} = Z_A^F f_\pi^F\right|_{\rm lat}.
\end{equation}
As a special case, we have
\begin{equation}
Z_A^5 = 1
\label{eq:ZA5}
\end{equation}
in the $\xi_5$ channel due to the lattice PCAC relation~(\ref{eq:PCAC}).
Thus the pion decay constant can be calculated with out any
uncertainties of renormalization in this channel, while the other
channels can be used to check the reliability of renormalization
constants by examining the expected convergence of the renormalized pion
decay constants to a single value in the continuum limit.

The decay constant defined with the pion operator~(\ref{eq:fpiP}) is
renormalized as
\begin{equation}
f_\pi^{(P)5}\big|_{\rm phys} = (Z_P^5 / Z_m) f_\pi^{(P)5}
\big|_{\rm lat},
\end{equation}
where $Z_m$ is the renormalization constant for quark mass.  Using the
identities $Z_m = 1 / Z_S^I$ and $Z_S^I = Z_P^5$, where the superscript
$I$ refers to the KS flavor for a unit matrix, we find that this
relation is identical to
\begin{equation}
f_\pi^{(P)5}\big|_{\rm phys} = f_\pi^{(P)5}\big|_{\rm lat},
\label{eq:fpi3}
\end{equation}
which is equivalent to Eq.~(\ref{eq:ZA5}).

\subsection{Perturbative and non-perturbative renormalization factors
            for axial vector currents}

We employ two sets of the renormalization factor $Z_A^F$ for the KS
axial vector current.  One of them is perturbatively calculated at one-%
loop order\cite{pertur}.  We apply tadpole improvement to the axial
vector current operator using the fourth root of plaquette as the
tadpole factor, and evaluate the renormalization constants with the
tadpole-improved $\overline{\rm MS}$ coupling at $q^* = 1 / a$.  The
other is non-perturbatively evaluated with the regularization
independent (RI) scheme of Ref.~\cite{Mart95}, which was developed for
the Wilson and clover actions.  In the RI scheme, the renormalization
factor is obtained from the amputated Green function in momentum space
\begin{equation}
\Gamma_{O_S}^F(p) = S(p)^{-1}\langle 0 |\phi(p) O_S^F
\overline{\phi}(p) | 0\rangle S(p)^{-1},
\label{eq:green}
\end{equation}
where the quark two-point function is defined by
$S(p) =\langle 0 |\phi(p)\overline{\phi}(p) | 0\rangle$, and the
momentum of the hypercubic field $\phi(p)$ takes values of the form
$p_\mu = 2\pi n_\mu / (a L)$ with $- L / 4\leq n_\mu\leq L / 4 - 1$.
The renormalization condition imposed upon $\Gamma_{O_S}^F(p)$ is given
by
\begin{equation}
Z_{O_S}^{{\rm (RI)}F}(p) Z_\phi(p) = {\rm Tr} [(
{\cal P}_{O_S}^F)^\dagger\Gamma_{O_S}^F(p)],
\end{equation}
where $\left({\cal P}_{O_S}^F\right)^\dagger =\big(\gdoxd{S}{F}\big)$ is
the projector onto the tree-level amputated Green function.  The wave
function renormalization constant $Z_\phi$ is calculated by imposing the
condition $Z_V^I(p) = 1$ for the conserved vector current for
$(\goI{\mu})$.  The relation between the overall renormalization
constant $Z_A^F$ appearing in Eq.~(\ref{eq:ZA}) and $Z_A^{{\rm (RI)}F}$
is simply
\begin{equation}
Z_A^F = 1 / Z_A^{{\rm (RI)}F},
\end{equation}
because the continuum axial vector current is not renormalized.

The calculations for the non-perturbative renormalization constants were
carried out in quenched QCD in our previous publication\cite{JLQC99}.
The results for the scalar and pseudoscalar operators have been used in
our analysis of light quark masses for the KS quark action in quenched
QCD\cite{JLQC99}.  Here we use them for the axial vector renormalization
factors.
  
\begin{figure}[p]
\begin{tabular}{lc}
\epsfxsize=222pt\epsfbox{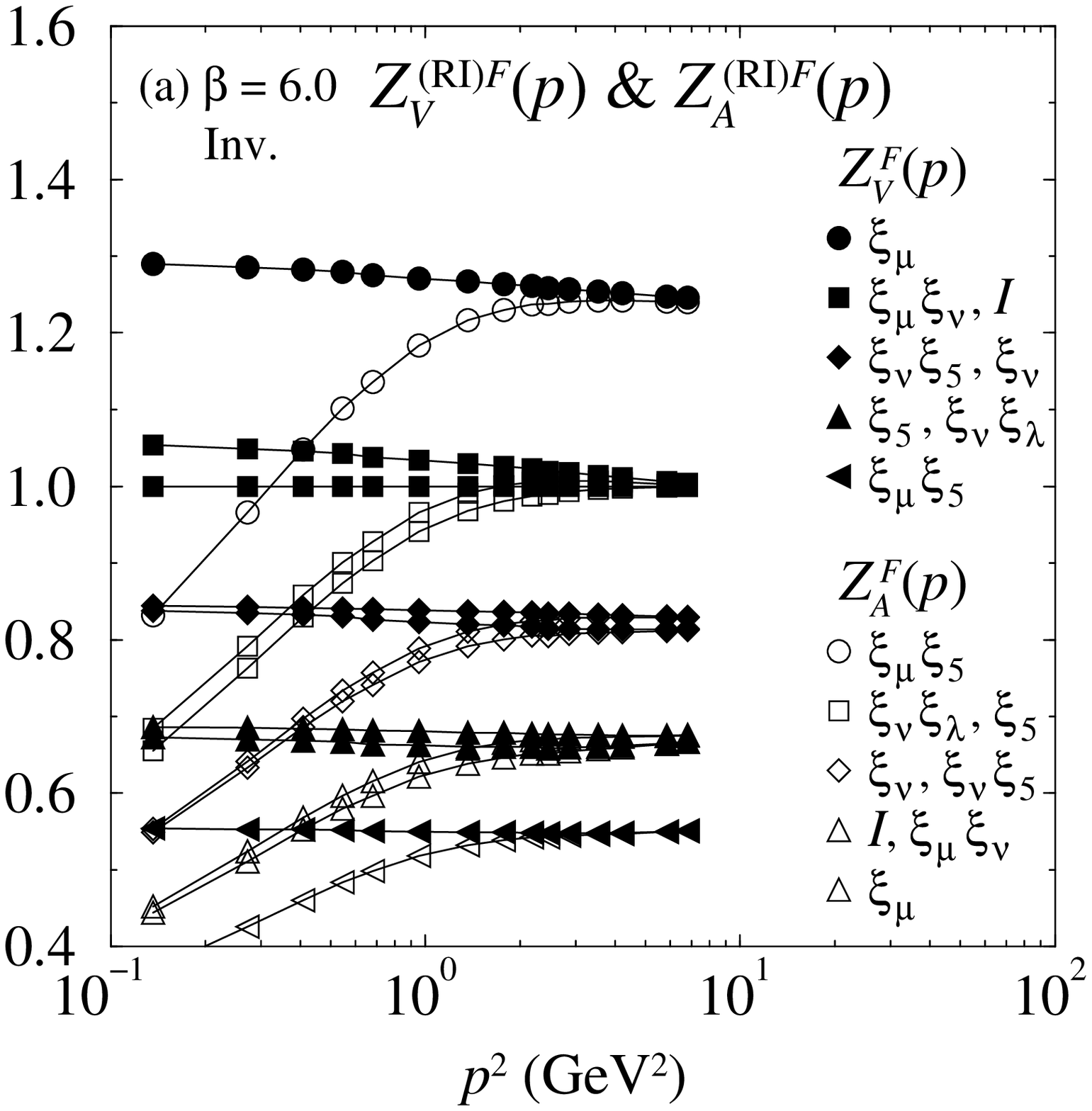} &
\epsfxsize=222pt\epsfbox{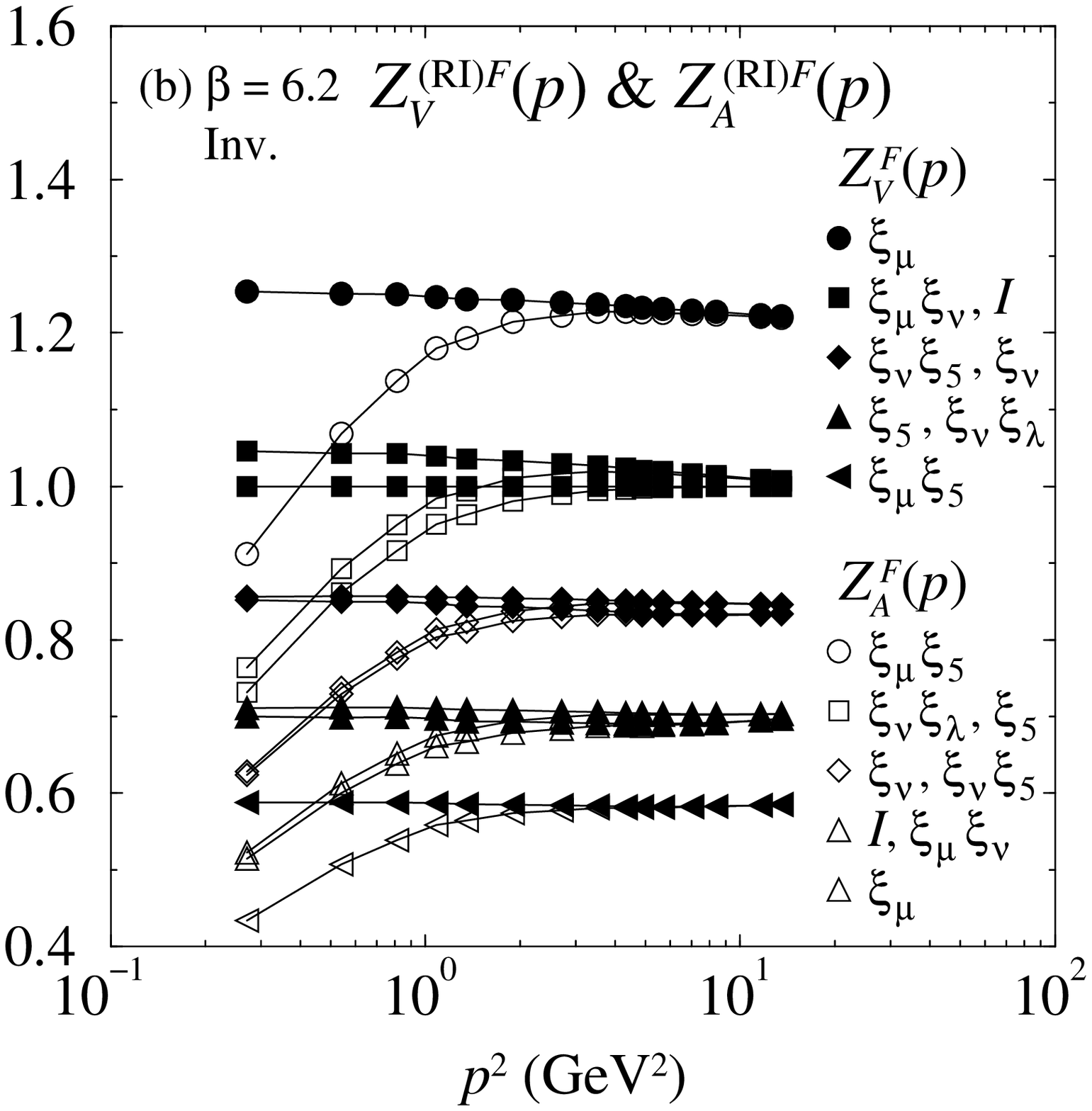} \\
\epsfxsize=222pt\epsfbox{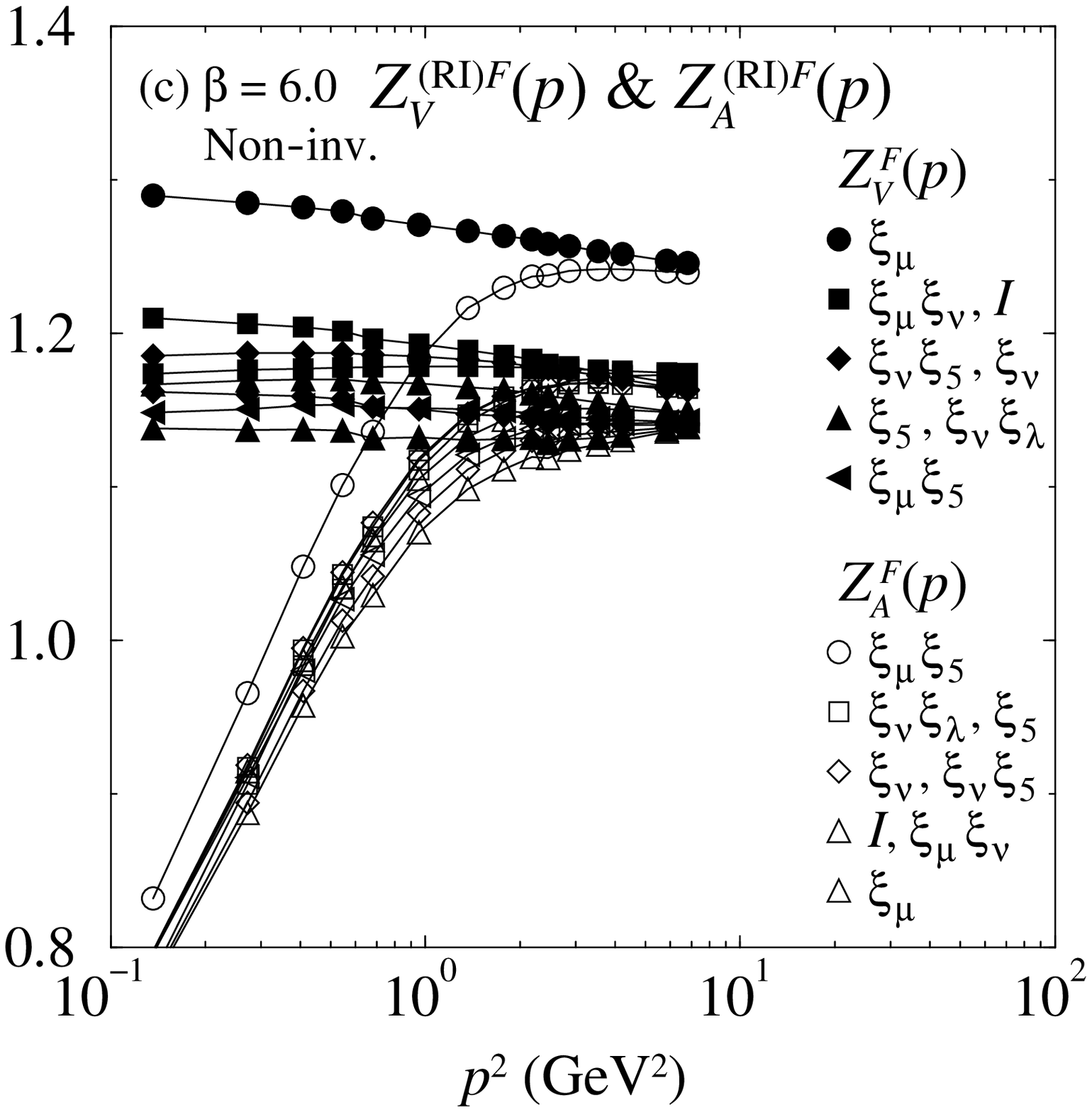} &
\epsfxsize=222pt\epsfbox{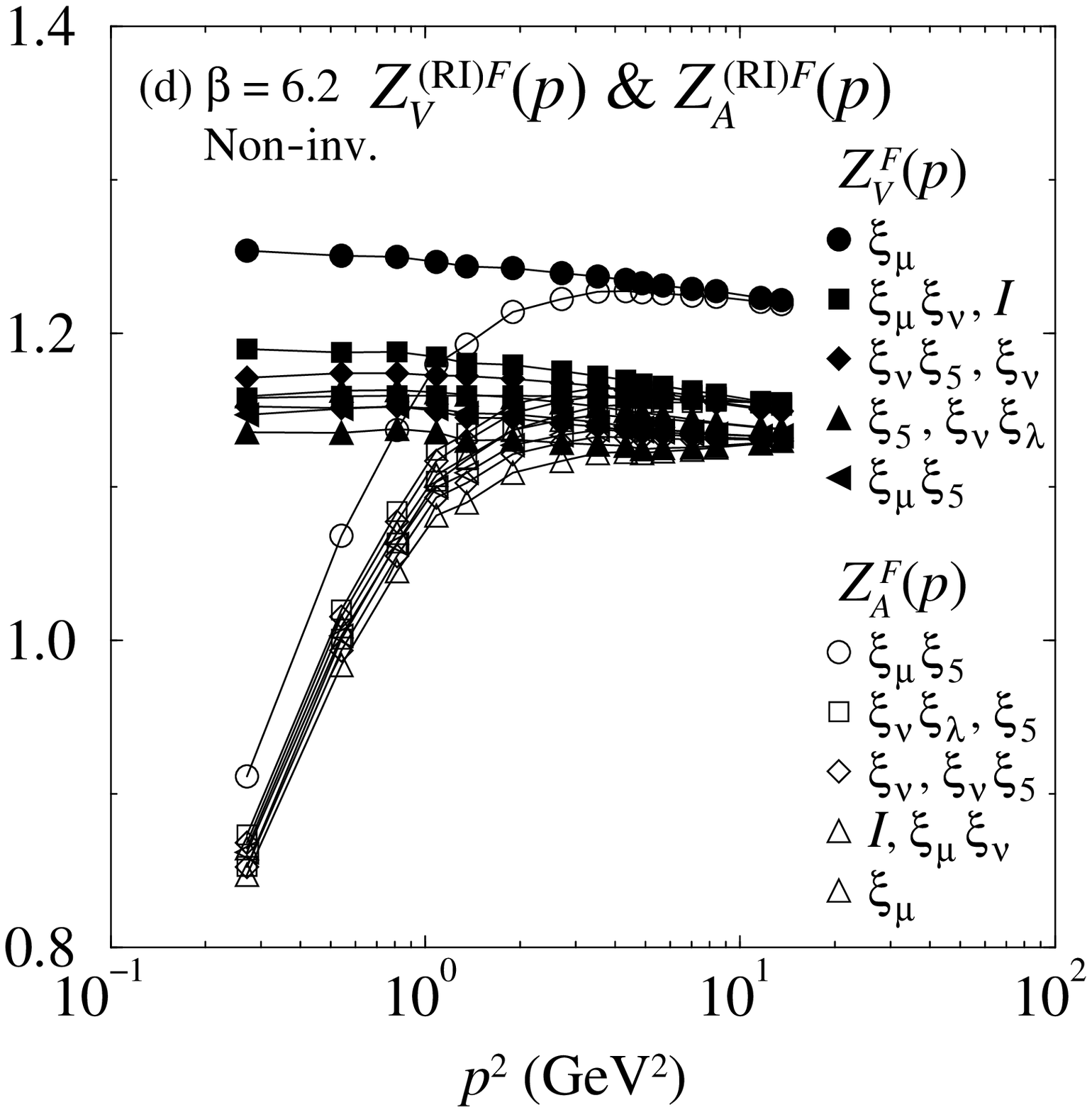} \\
\end{tabular}
\caption{Non-perturbative renormalization constants for vector
         $Z_V^F(p)$ and axial vector currents $Z_A^F(p)$ examined
         with gauge invariant current for (a) $\beta = 6.0$ and (b)
         $\beta = 6.2$, and by gauge non-invariant current for (c)
         $\beta = 6.0$ and (d) $\beta = 6.2$.}
\label{fig:Zs}
\end{figure}

The calculational parameters are summarized in Table~\ref{tab:parameter}.
We evaluate the Green function~(\ref{eq:green}) for 15 momenta in the
range $0.038533\leq (p a)^2\leq 1.9277$ using quark propagators
evaluated with a source in a momentum eigenstate.  In Fig.~\ref{fig:Zs}
we present the renormalization constant for both vector and axial vector
currents, respectively denoted by $Z_V^{{\rm (RI)}F}(p)$ and
$Z_A^{{\rm (RI)}F}(p)$, in the chiral limit.

A practically important issue with the non-perturbative method employed
here is the choice of the momentum at which the renormalization factors
are evaluated.  In general the momentum should satisfy
$\Lambda_{\rm QCD}\ll p\ll O(a^{-1})$ in order to keep under control the
non-perturbative hadronization effects and the discretization error on
the lattice.  Since these effects appear as $p$-dependences of
renormalization factors, we should avoid the range where a momentum
dependence is visible.  Another point to consider is the relation
$Z_V^{{\rm (RI)}F5}(p) = Z_A^{{\rm (RI)}F}(p)$ with the superscript $F5$
referring $\xi_F\xi_5$, which we would expect to hold for all momenta
$p$ in the chiral limit due to $\rm U(1)_A$ chiral symmetry of the KS
quark action.

For $\beta = 6.2$ Fig.~\ref{fig:Zs} shows that these two requirements
are satisfied for $p^2 > 5$ GeV$^2$, which corresponds to
$(p a)^2 > 0.5$.  In order to satisfy $p\ll O(a^{-1})$, we take
$(p a)^2 = 1.0024$ ($p^2 = 7.0392$ GeV$^2$ in physical units) to
calculate the renormalization factors used for the pion decay constant.
The same value of lattice momentum $(p a)^2 = 1.0024$ is chosen for
$\beta = 6.0$, which corresponds to $p^2 = 3.5428$ GeV$^2$.  The
numerical values of the renormalization factors are summarized in Table~%
\ref{tab:Zs}.

\section{Details of simulation}
\label{sec:simulation}

\subsection{Simulation parameters}

We carry out our calculations in quenched QCD using the standard
plaquette action for gluons.  As we summarize in Table~%
\ref{tab:parameter2}, numerical simulations are carried out at
$\beta\equiv 6 / g^2 = 6.0$ and 6.2 on $32^3\times 64$ and
$48^3\times 64$ lattices, respectively.  Gauge configurations are
generated with the five-hit pseudo-heatbath algorithm, and hadron
correlation functions are calculated on 100(60) configurations separated
by 2000 sweeps at $\beta = 6.0$(6.2).

Gauge configurations are fixed to the Landau gauge through maximization
of
\begin{equation}
F_L =\sum_{n,\mu} {\rm tr} (U_\mu(n) + U^\dagger_\mu(n)).
\end{equation}
This is realized by iterating the steepest descent method for the first
2000 steps and the over-relaxation method for the subsequent 3000 steps
until the condition
\begin{equation}
\Delta =\frac{1}{6 V}\sum_n {\rm tr} (G_L^\dagger(n) G_L(n)) <
10^{-14}
\end{equation}
is satisfied, where $V$ is the lattice volume and
\begin{equation}
G_L =\sum_\mu\frac{1}{2} (U_\mu(n) - U_\mu(n -\hat{\mu}) -\mbox{h.c.} -
\mbox{trace}).
\end{equation}

We take three values for quark mass, $m_q a = 0.030$, 0.020, 0.010 at
$\beta = 6.0$ and 0.023, 0.015, 0.008 at $\beta = 6.2$.  Quark
propagators are evaluated for 16 types of wall sources, each
corresponding to a corner of a hypercube, defined by
\begin{equation}
\sum_{y,B} D_{AB}(x, y)\sum_{\vec{z}} G_{BC}(y, z) =\sum_{\vec{z}}
\delta_{xz}\delta_{AC},
\end{equation}
where $D_{AB}(x, y)$ is the quark matrix for the KS action.  We solve
the equation independently for each $C$ by the conjugate gradient method
with the stopping condition
\begin{equation}
||\mbox{remnant vector}||^2 < 10^{-5}.
\end{equation}
The 16 quark propagators are combined to construct the 16 meson
correlation functions in the KS flavor basis specified by the hypercube
matrix $\xi_F$.  Averages are taken of the meson correlation functions
over $2^3$ ways of choosing the spatial origin of hypercubes on the
lattice.  We also average them over all states belonging to the same
irreducible representation\cite{Golt86}.

\subsection{Fitting procedure}

In fitting the meson correlation function $C(t)$ to the asymptotic form
$C^{\rm fit}(t)$ for an extraction of the mass and amplitude, we
symmetrize the correlator at $t$ and $T-t$, and carry out a standard
correlated fit minimizing
\begin{equation}
\chi^2 =\sum_{t,t'}\Delta C(t)\Sigma^{-1}(t, t')\Delta C(t'),
\end{equation}
where
\begin{equation}
\Sigma(t,t') =\langle C(t) C(t')\rangle -\langle C(t)\rangle\langle
C(t')\rangle
\end{equation}
is the covariance matrix of the correlator and
$\Delta C(t) = C(t) - C^{\rm fit}(t)$.  The fitting range
$t = t_{\rm min},\ldots, t_{\rm max}$ is chosen by fixing
$t_{\rm max} = T/2$ and varying $t_{\rm min}$ so that
$\chi^2 / N_{\rm DF}$ takes a value near unity, where $N_{\rm DF}$ is
the degree of freedom of the fit.  Finally, errors in this work are
estimated by the single elimination jackknife procedure.

\subsection{Wall-to-wall amplitude}

\begin{figure}[p]
\begin{tabular}{lc}
\epsfxsize=222pt\epsfbox{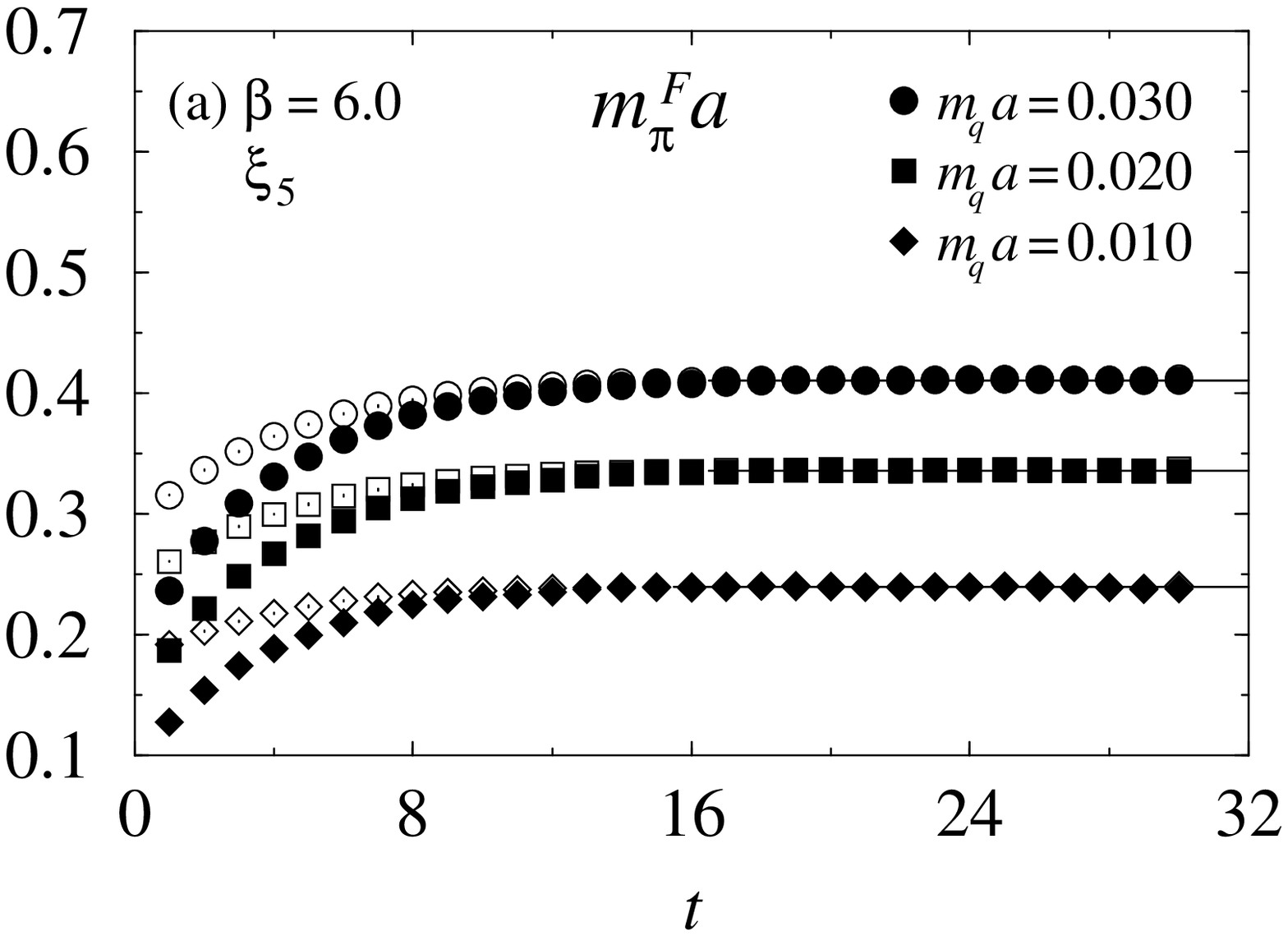} &
\epsfxsize=222pt\epsfbox{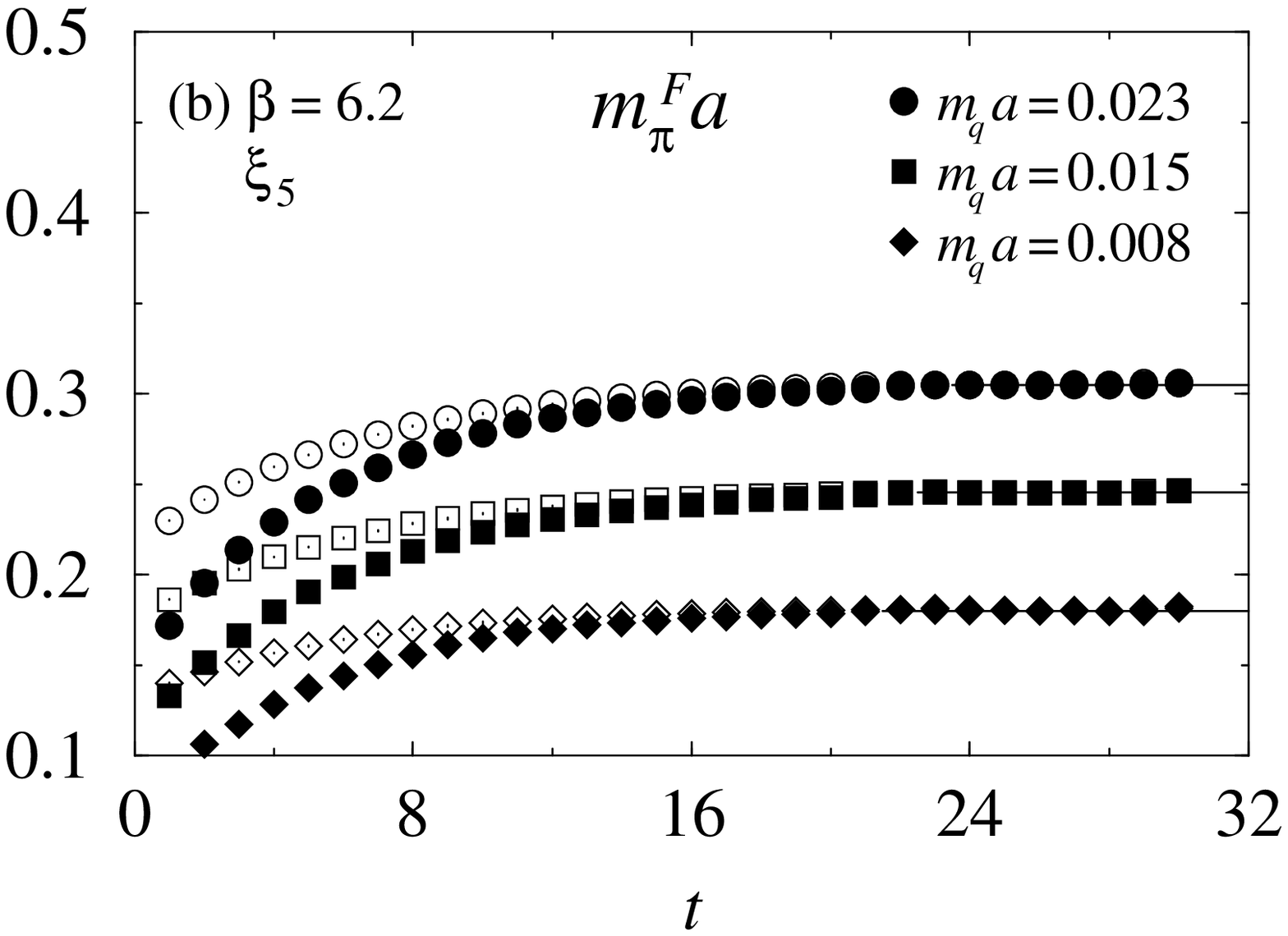} \\
\end{tabular}
\caption{Typical comparison of the effective mass of axial vector
         current (open symbols) and pion (filled symbols) correlation
         functions with its global mass (horizontal lines) for
         $\xi_F = \xi_5$ at (a) $\beta = 6.0$ and (b) $\beta = 6.2$.
         Circles, squares and diamonds refer to the quark masses in
         descending order at each coupling respectively.  Note that
         result does not depend on gauge invariance of the operator
         in the case using local operator like this case.}
\label{fig:effPA}
\end{figure}

We check the validity of the asymptotic form of the mesonic correlation
function~(\ref{eq:correlator}) which is based on the assumption of a
single pole dominance by an inspection of the effective mass.  Typical
results for the effective mass extracted from the correlators
$\langle\pi^F(t)\pi_W^F(0)\rangle$ and
$\langle A_4^F(t)\pi_W^F(0)\rangle$ are compared in Fig.~\ref{fig:effPA}.
We observe a wide plateau and an expected agreement of the effective
masses from the two correlation functions.  We then find no problem in
fitting these correlation functions by a single pole.

\begin{figure}[p]
\begin{tabular}{lc}
\epsfxsize=222pt\epsfbox{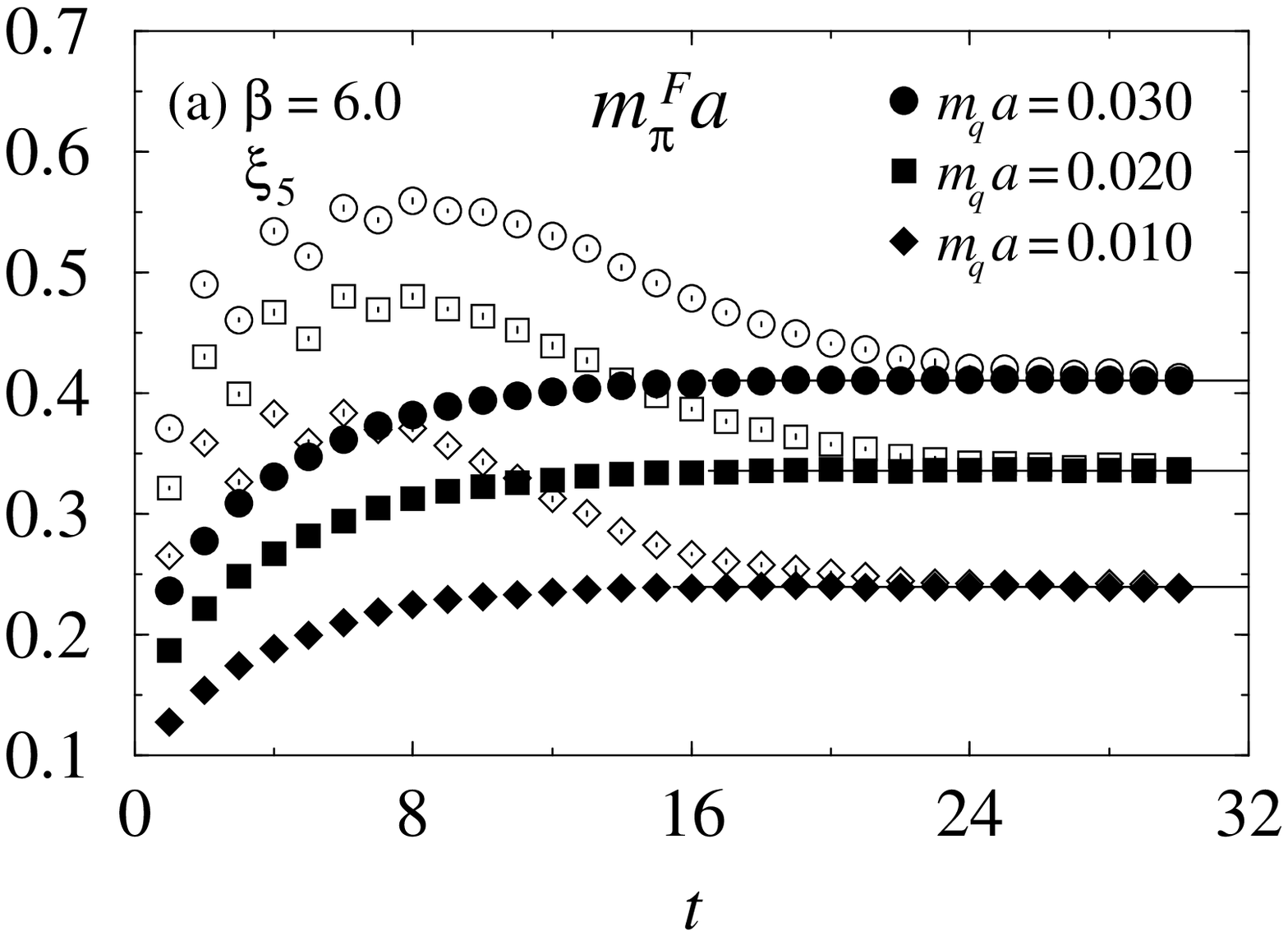} &
\epsfxsize=222pt\epsfbox{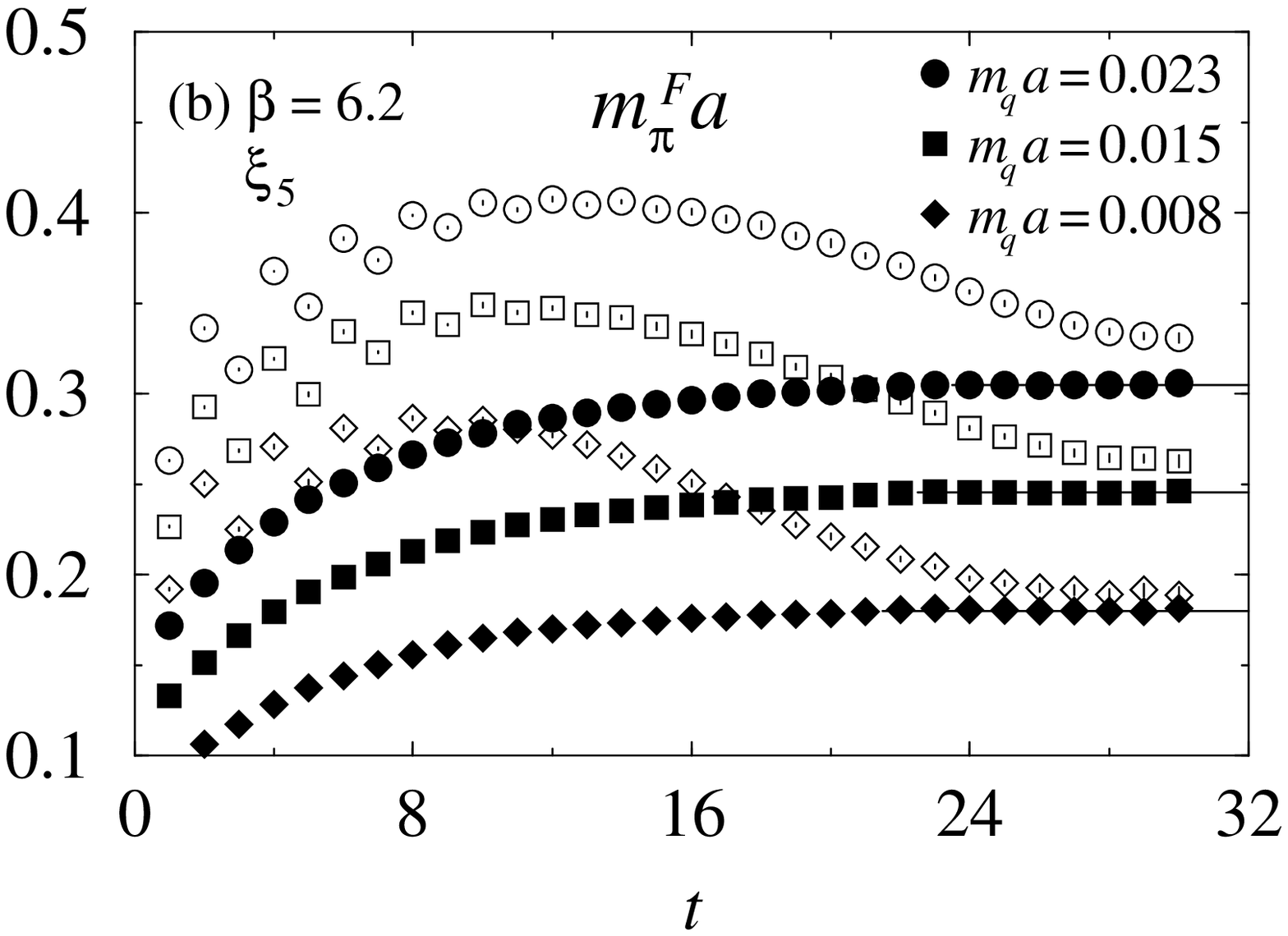} \\
\end{tabular}
\caption{Typical comparison of the effective masses of the wall-to-%
         wall correlation function (open symbols) and that for pion
         correlation function (filled symbols) with its global mass
         (horizontal lines) for $\xi_F = \xi_5$ at (a) $\beta = 6.0$
         and (b) $\beta = 6.2$.}
\label{fig:effPW}
\end{figure}

The situation is different for the wall-to-wall correlation function
$\langle\pi_W^F(t)\pi_W^F(0)\rangle$, particularly at $\beta = 6.2$.  As
we show in Fig.~\ref{fig:effPW}, the effective mass for
$\langle\pi_W^F(t)\pi_W^F(0)\rangle$ does not reach a plateau at
$\beta = 6.2$ even at $t\sim T/2$, and agreement with the effective mass
of $\langle\pi^F(t)\pi_W^F(0)\rangle$ is not seen.  This behavior is
most likely caused by a lack of sufficient temporal size of the lattice,
and poses a practical problem of how one extracts the wall-to-wall
amplitude $C_{\pi_W\pi_W}^F$ which is needed in Eq.~(\ref{eq:fpi2}) to
calculate the pion decay constant.

To solve this problem, we perform a double pole fit for
$\langle\pi_W^F(t)\pi_W^F(0)\rangle$ given by
\begin{eqnarray}
\makebox[10pt][l]{$\langle\pi_W^F(t)\pi_W^F(0)\rangle$} &
 & \nonumber\\
&&\sim C_{\pi_W\pi_W}^F (\sigma_t)^t [\exp(- m_\pi^F\,t) +\exp(-
   m_\pi^F\,(T - t))] \nonumber\\
&&\;+\;C_{q\overline{q}}^F\:\,[\exp(- m_{q\overline{q}}^F t) +\exp(-
   m_{q\overline{q}}^F(T - t))].
\end{eqnarray}
Ideally one likes to make a fit with four parameters $C_{\pi_W\pi_W}^F$,
$m_\pi^F$, $C_{q\overline{q}}^F$, and $m_{q\overline{q}}^F$.  This fit,
however, is quite unstable because the fitting function consists of a
sum of two exponentials with not much different masses $m_\pi^F$ and
$m_{q\overline{q}}^F$.  Therefore, we fix the pion mass parameter
$m_\pi^F$ to that obtained from $\langle\pi^F(t)\pi_W^F(0)\rangle$.

\begin{figure}[p]
\begin{tabular}{lc}
\epsfxsize=222pt\epsfbox{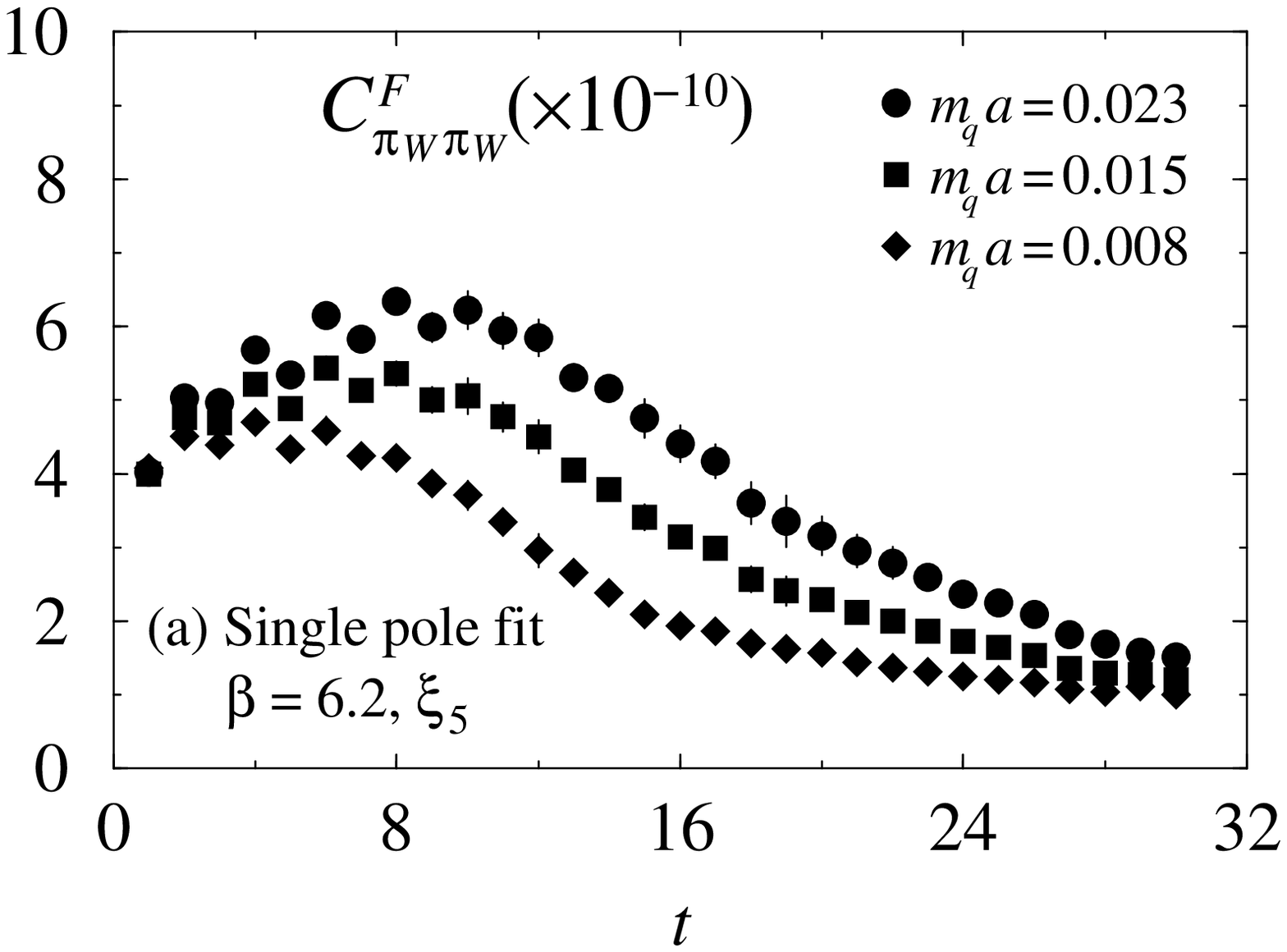} &
\epsfxsize=222pt\epsfbox{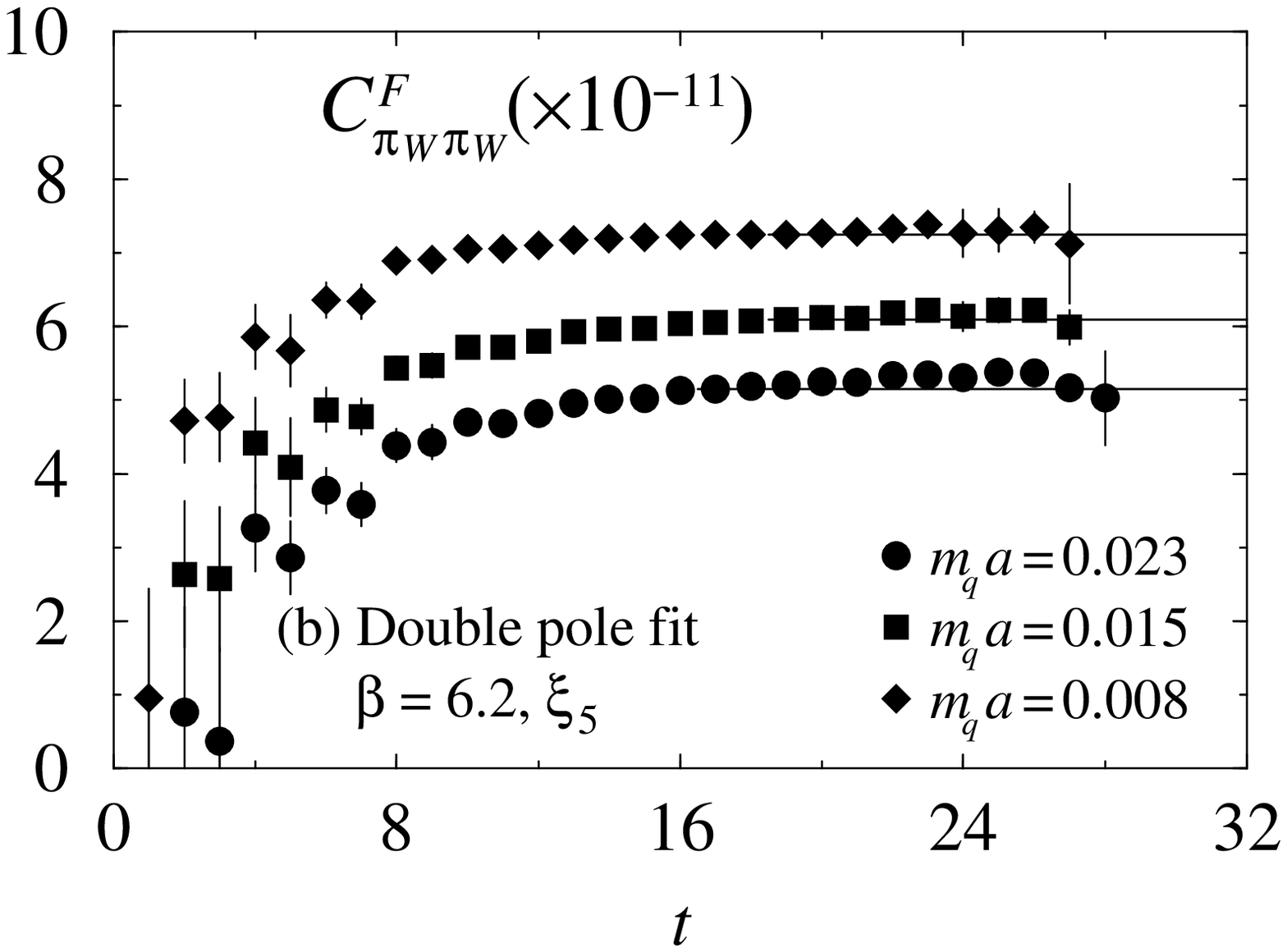} \\
\end{tabular}
\caption{Typical comparison of of the amplitude of wall-to-wall pion
         correlation function for $\xi_F = \xi_5$ at $\beta = 6.2$
         obtained by (a) the single pole fit and (b) the double pole
         fit.}
\label{fig:effAmp}
\end{figure}

\begin{figure}[p]
\begin{tabular}{lc}
\epsfxsize=222pt\epsfbox{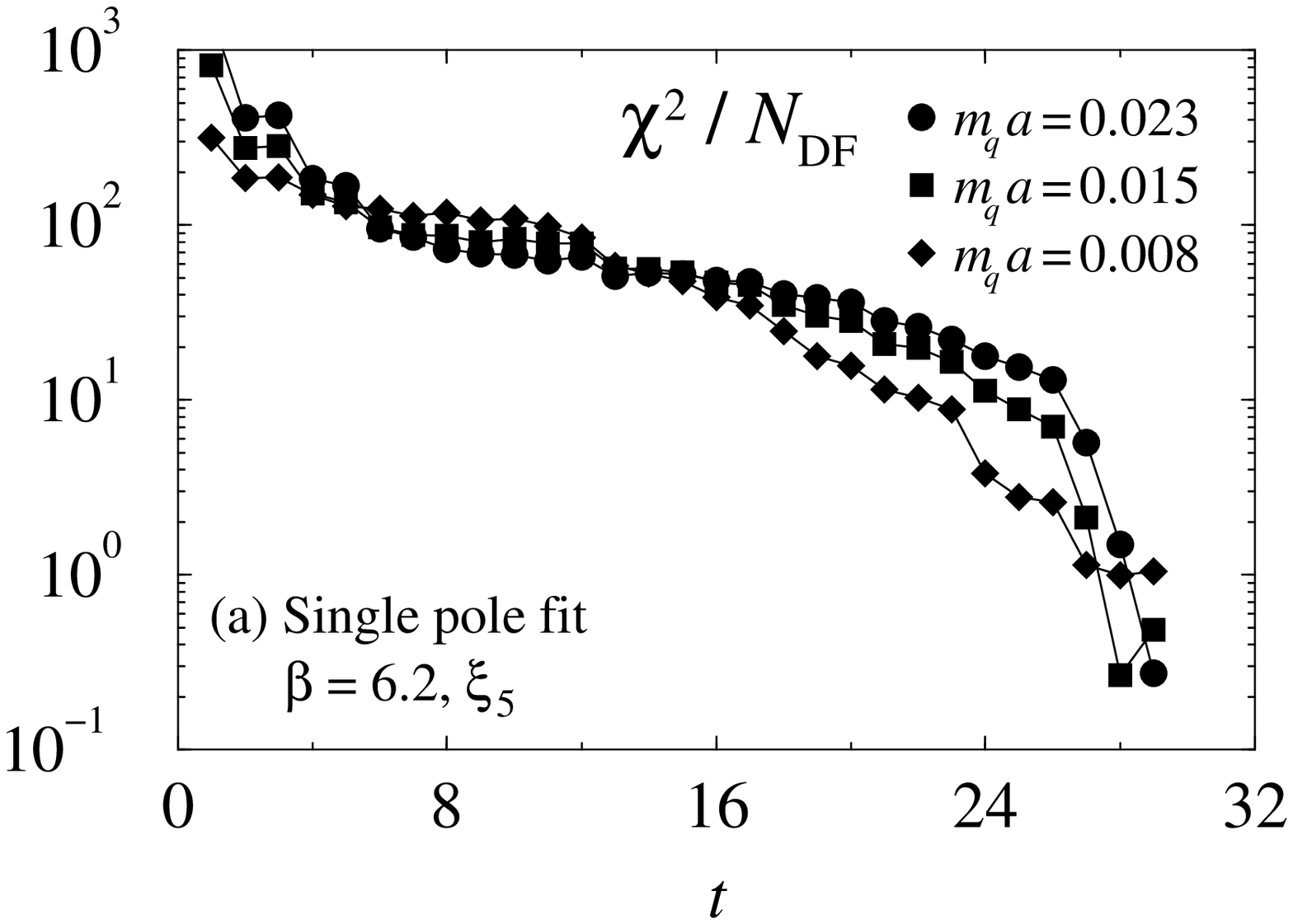} &
\epsfxsize=222pt\epsfbox{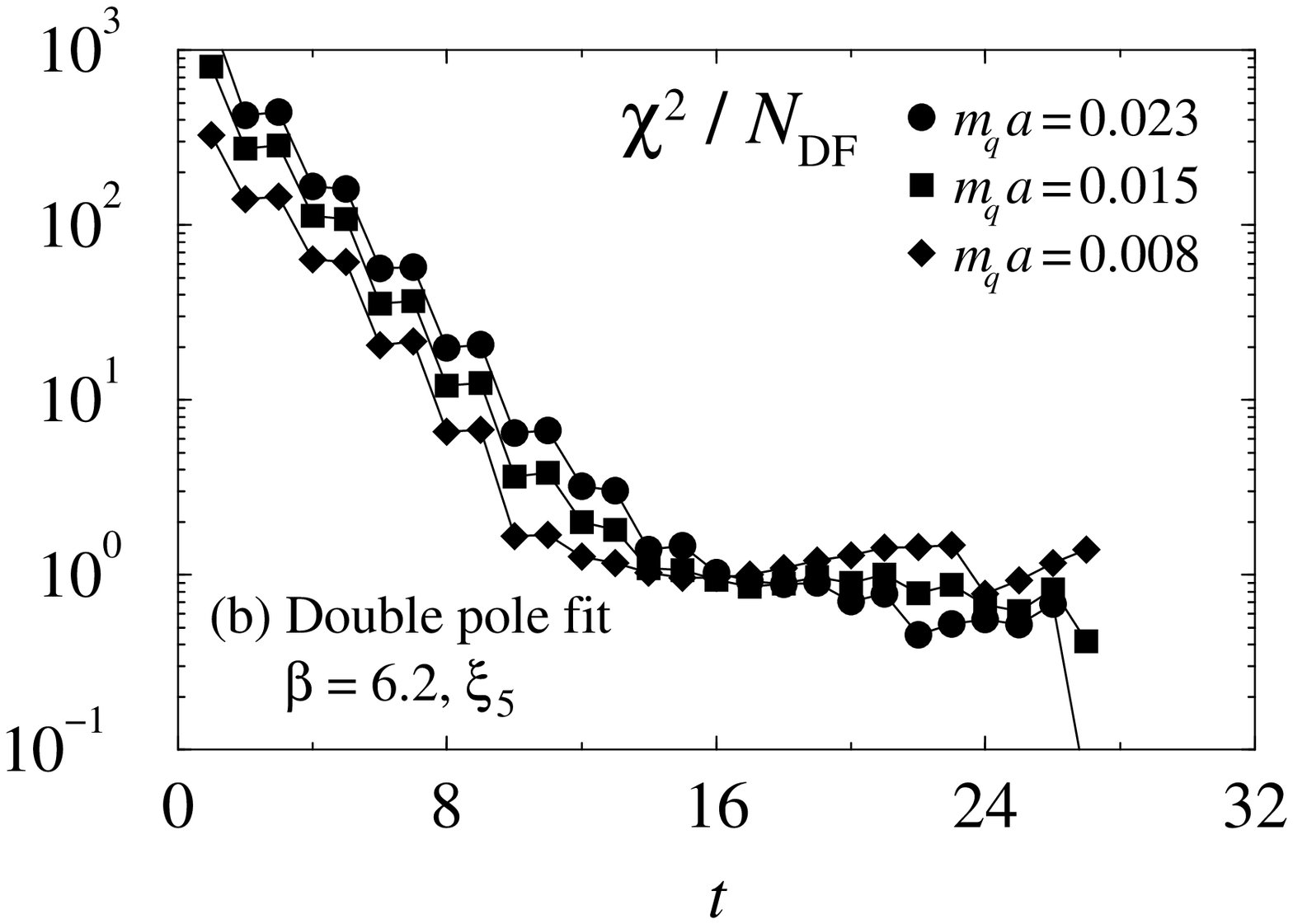} \\
\end{tabular}
\caption{Typical comparison of $\chi^2 / N_{\rm DF}$ for (a) the
         single pole fitting and (b) the double pole fitting of wall-%
         to-wall pion correlation function for $\xi_F = \xi_5$ at
         $\beta = 6.2$.}
\label{fig:chisq}
\end{figure}

As we now can no longer compare the effective pion mass for
$\langle\pi_W^F(t)\pi_W^F(0)\rangle$ to that for
$\langle\pi^F(t)\pi^F(0)\rangle$, we present a typical comparison of the
amplitudes, extracted with the fitting range from $t$ to $T / 2$ with
the single and double pole fits, in Fig.~\ref{fig:effAmp}.  We also
compare $\chi^2 / N_{\rm DF}$ for the two fits in Fig.~\ref{fig:chisq}.
From these figures, we consider that the double pole fit provides a good
determination of the amplitude $C_{\pi_W\pi_W}^F$ of the pion to the
wall operator with a wide plateau of the amplitude and a reasonable
value of $\chi^2 / N_{\rm DF}\sim O(1)$.

\begin{figure}[p]
\begin{tabular}{lc}
\epsfxsize=222pt\epsfbox{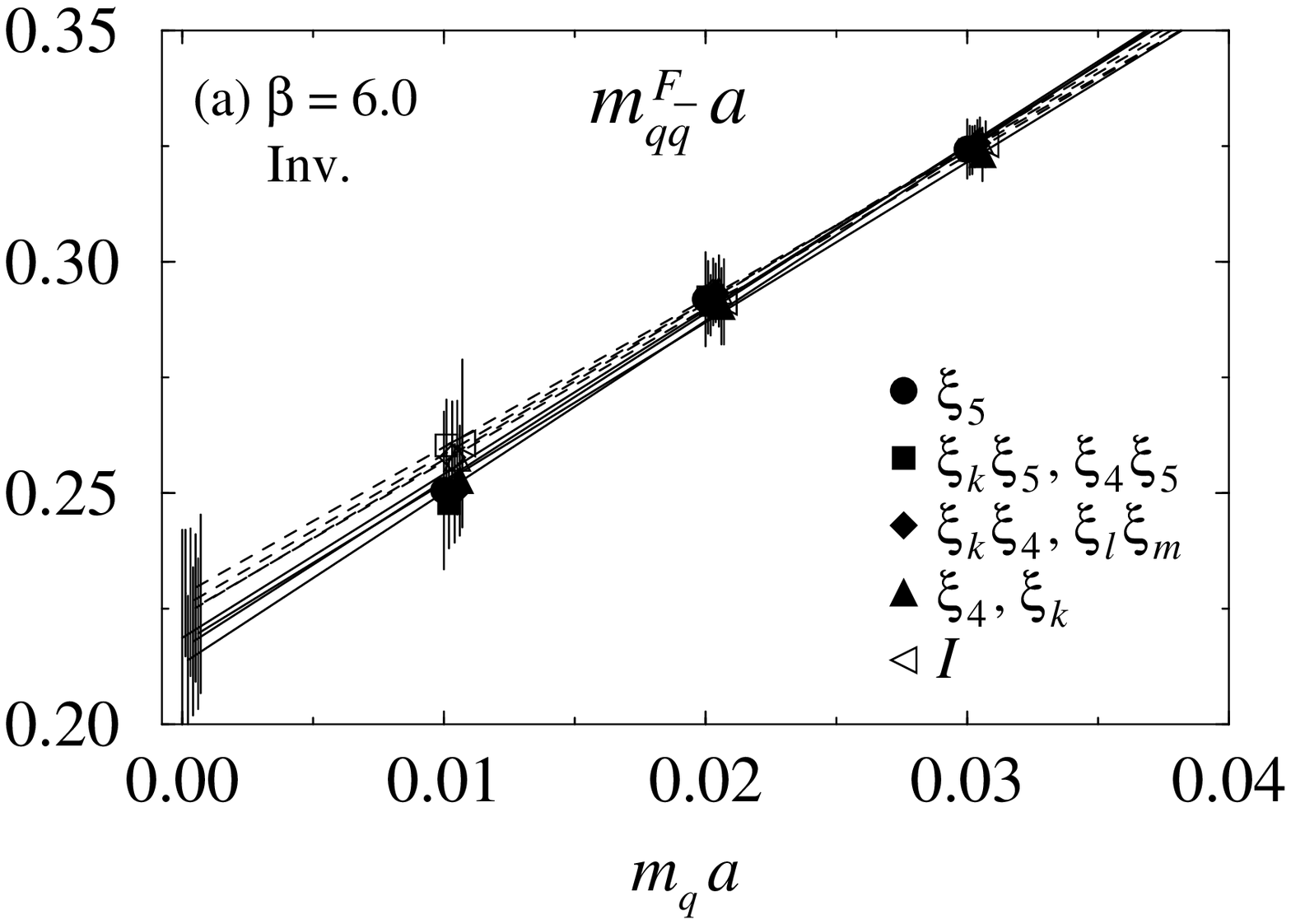} &
\epsfxsize=222pt\epsfbox{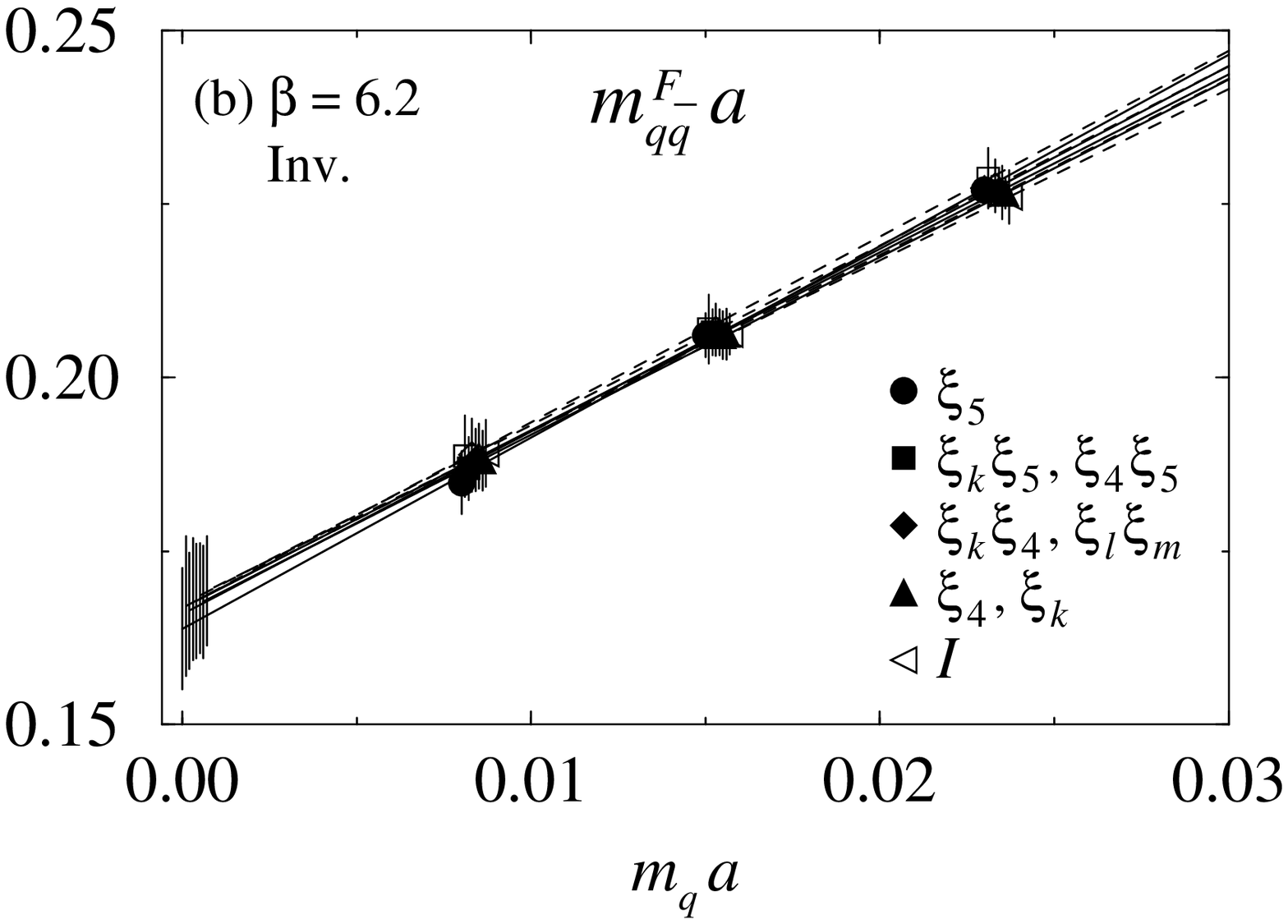} \\
\end{tabular}
\caption{Chiral behavior of the alternative pole masses
         appearing in the wall-to-wall correlation function
         at (a) $\beta = 6.0$ and (b) $\beta = 6.2$.
         Shape of symbols refer to the distance of the operator.
         Some symbols (square, diamond and up-triangle) denotes two
         flavors; the former one refers time-local operators (filled
         symbols) including flavor $\xi_5$, and the latter one refers
         time-separated operators (open symbols).}
\label{fig:massQ}
\end{figure}

A possible interpretation for the dominant source of contamination to
the wall-to-wall correlation function is an unbound quark-antiquark pair.
Such a state can contribute since gauge configurations are fixed to the
Landau gauge.  In Fig.~\ref{fig:massQ} we plot the value of the second
pole mass $m_{q\overline{q}}$ as a function of quark mass.  The fact
that the results depend little on the KS flavor of the meson operator is
consistent with this interpretation. In the chiral limit one obtains
$m_{q\overline{q}}\sim 2\times 440$ MeV, which is a reasonable value for
a constituent quark mass.

Finally, we summarize the fitting ranges $t_{\rm min}$ common for all
flavors and $\chi^2 / N_{\rm DF}$ for our global fits in Table~%
\ref{tab:range}.  Here, we have used the alternative fitting range of
the wall-to-wall correlation function to improve the fitting quality for
$\xi_F = \xi_4$, because the common fitting range does not give a
satisfactory result\cite{Kane99} caused by worse fitting.

\section{Chiral behavior}
\label{sec:chiral}

\subsection{Pion masses}

\begin{figure}[p]
\begin{tabular}{lc}
\epsfxsize=222pt\epsfbox{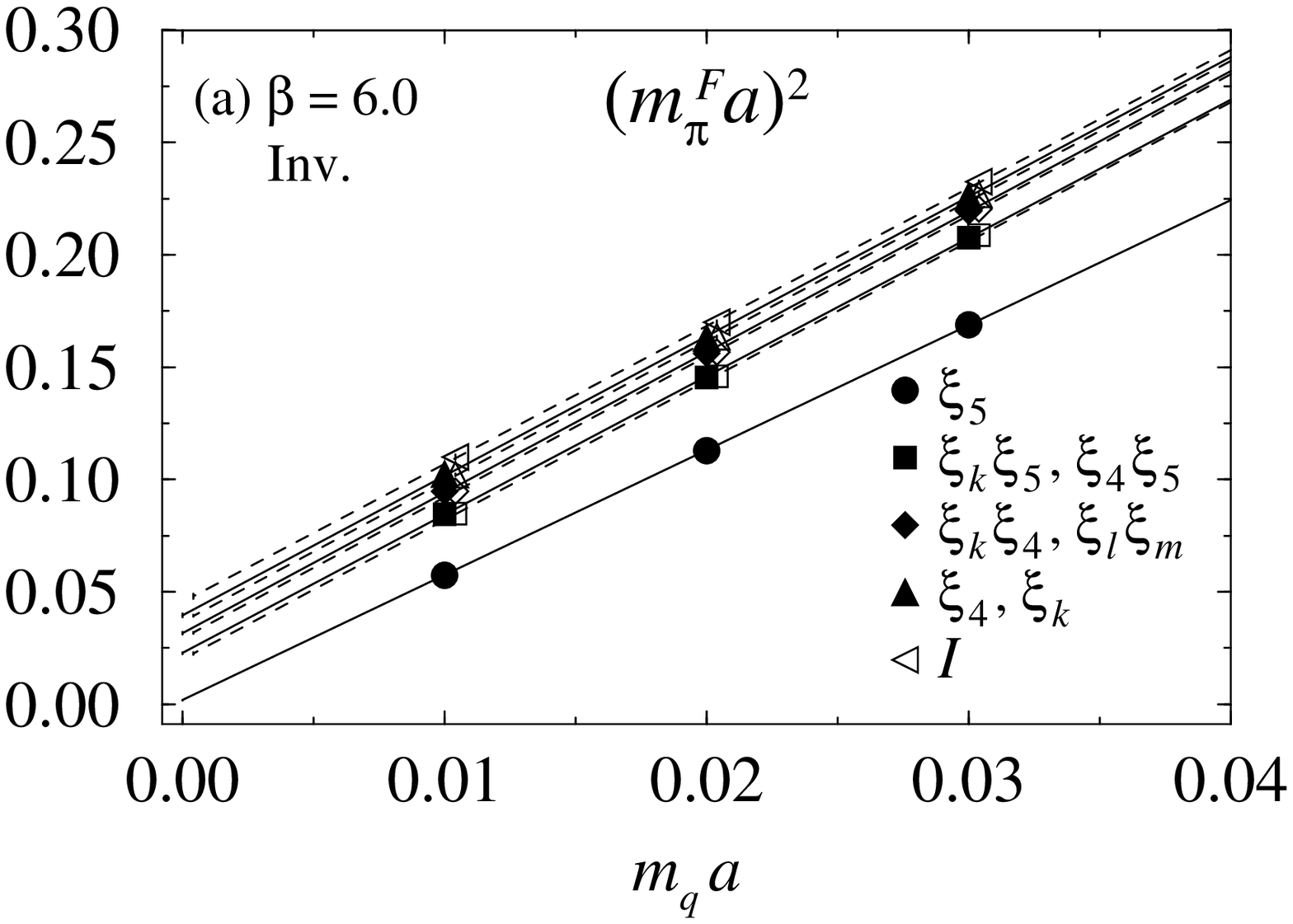} &
\epsfxsize=222pt\epsfbox{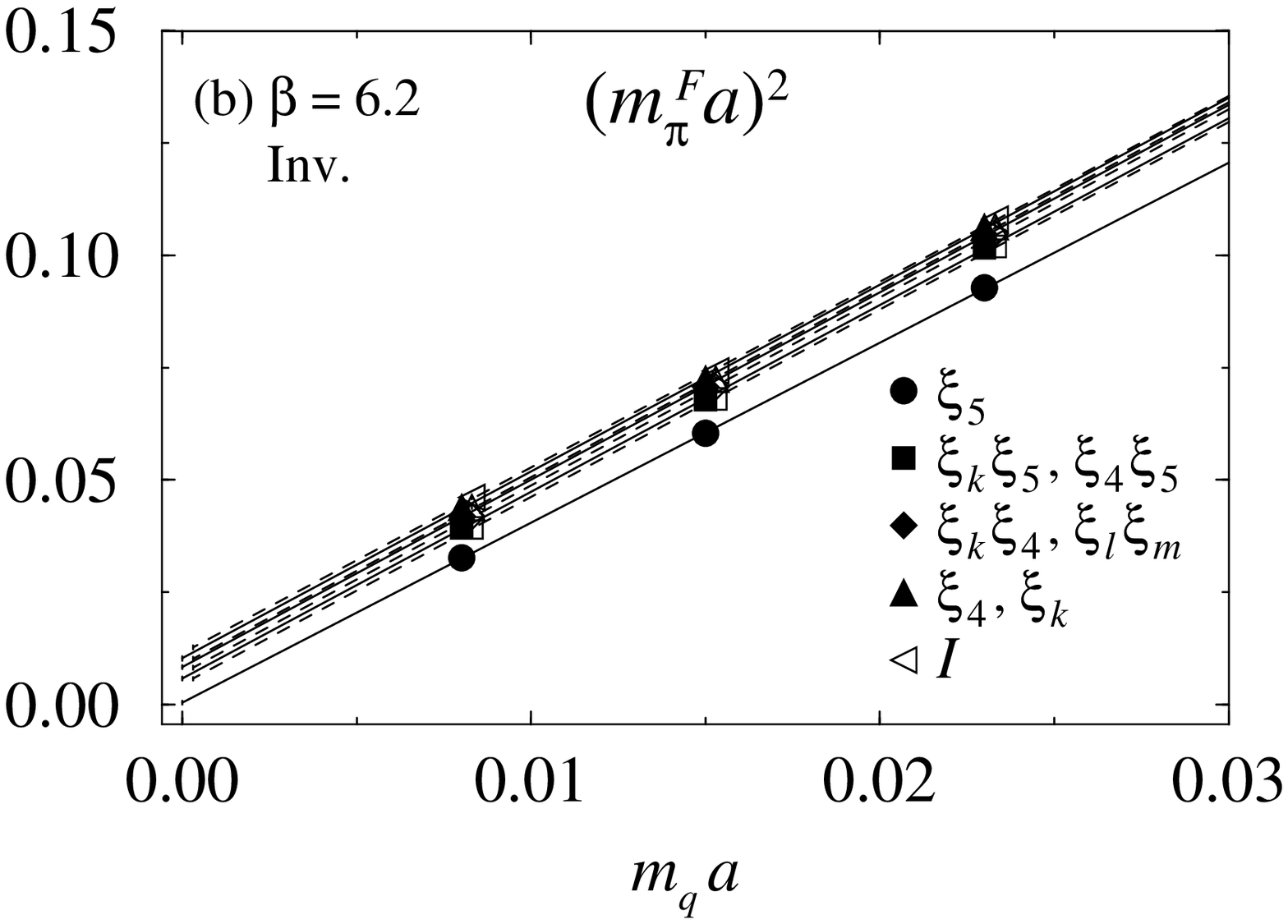} \\
\end{tabular}
\caption{Chiral behavior of pion masses obtained with the gauge
         invariant pion operators at (a) $\beta = 6.0$ and (b)
         $\beta = 6.2$.  Shape of symbols refer to the distance of
         the operator.  Some symbols denotes two flavors; the former
         one refers time-local operators (filled symbols) including
         flavor $\xi_5$, and the latter one refers time-separated
         operators (open symbols).  For gauge non-invariant result,
         see Table~\ref{tab:mass1}.}
\label{fig:mass1}
\end{figure}

We show values of $(m_\pi^F a)^2$ as a function of $m_q a$ in Fig.~%
\ref{fig:mass1}.  Pions for the 16 KS flavors are classified into 8
irreducible representations.  These consist of four 1-dimensional
representations given by $\xi_5$, $\xi_4\xi_5$, $\xi_4$, $I$ and four
3-dimensional representations given by $\xi_k\xi_5$, $\xi_k\xi_4$,
$\xi_k\xi_\ell$, $\xi_k$ $(k, \ell = 1, 2, 3; k < \ell)$.  We observe
very clearly in Fig.~\ref{fig:mass1} that these irreducible
representations form a degeneracy pattern specified by
\begin{equation}
\xi_5, (\xi_k\xi_5, \xi_4\xi_5), (\xi_k\xi_4, \xi_k\xi_\ell), (\xi_4,
\xi_k), I.
\end{equation}
This pattern was observed long time ago in Ref.~\cite{Ishi94}.  A
theoretical explanation based on the effective chiral Lagrangian
analysis for KS quark action was provided recently in Ref.~\cite{LeeS99}.

Another notable feature in Fig.~\ref{fig:mass1} is a linear behavior of
pion masses as a function of quark mass from the correlation function
with the gauge invariant pion operator.  With a linear extrapolation we
observe a non-vanishing value at $m_q a = 0$ in channels other than
$\xi_5$ for which $\rm U(1)_A$ symmetry holds.  The gauge non-invariant
case, not presented in the figure but in Table~\ref{tab:mass1} for the
numerical values, also shows almost the same result as in Fig.~%
\ref{fig:mass1}.

\begin{figure}[p]
\begin{tabular}{lc}
\epsfxsize=222pt\epsfbox{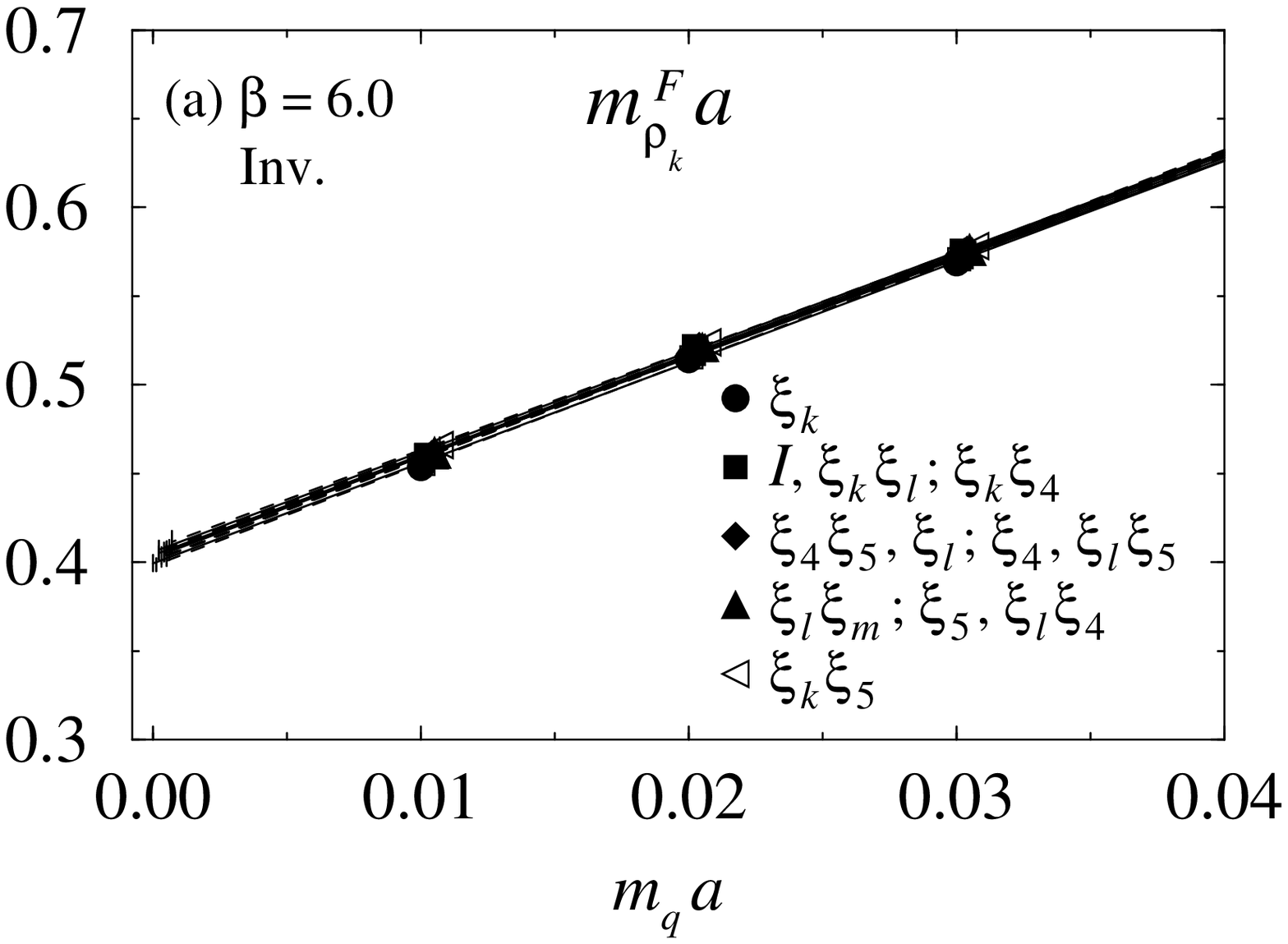} &
\epsfxsize=222pt\epsfbox{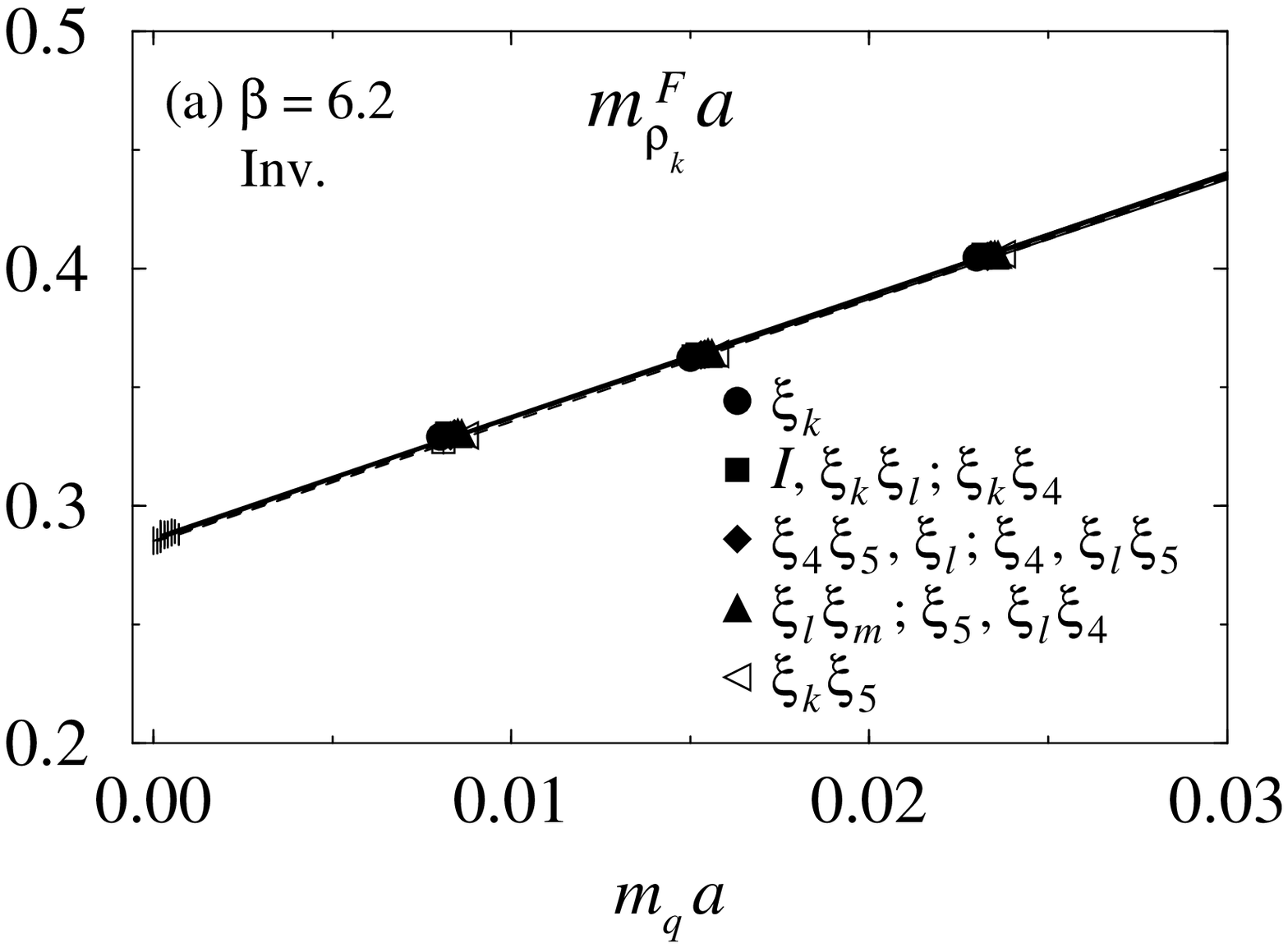} \\
\end{tabular}
\caption{Chiral behavior of gauge-invariant $\rho$ meson masses
         at (a) $\beta = 6.0$ and (b) $\beta = 6.2$.
         Symbols refer to the distance of the operator,
         Some symbols denotes four flavors;
         the first two flavors refer time-local operators (filled
         symbols), and the last two flavors refer time-separated
         operators (open symbols).}
\label{fig:massV}
\end{figure}

The chiral behavior of $\rho$ meson mass for various KS flavors is shown
in Fig.~\ref{fig:massV}.  We find the difference of masses among various
flavor channels to be small, less than 1\% even in the chiral limit
obtained by a linear extrapolation.  We therefore choose the $\rho$
meson mass in the flavor channel $(\gox{k}{k})$, for which the $\rho$
meson operator is local, to set the scale using the experimental value
$m_\rho = 770$ MeV.  We then find that $a^{-1} = 1.92(2)$ GeV for
$\beta = 6.0$ and $a^{-1} = 2.70(5)$ GeV for $\beta = 6.2$.

\subsection{Pion decay constant}

\begin{figure}[p]
\begin{tabular}{lc}
\epsfxsize=222pt\epsfbox{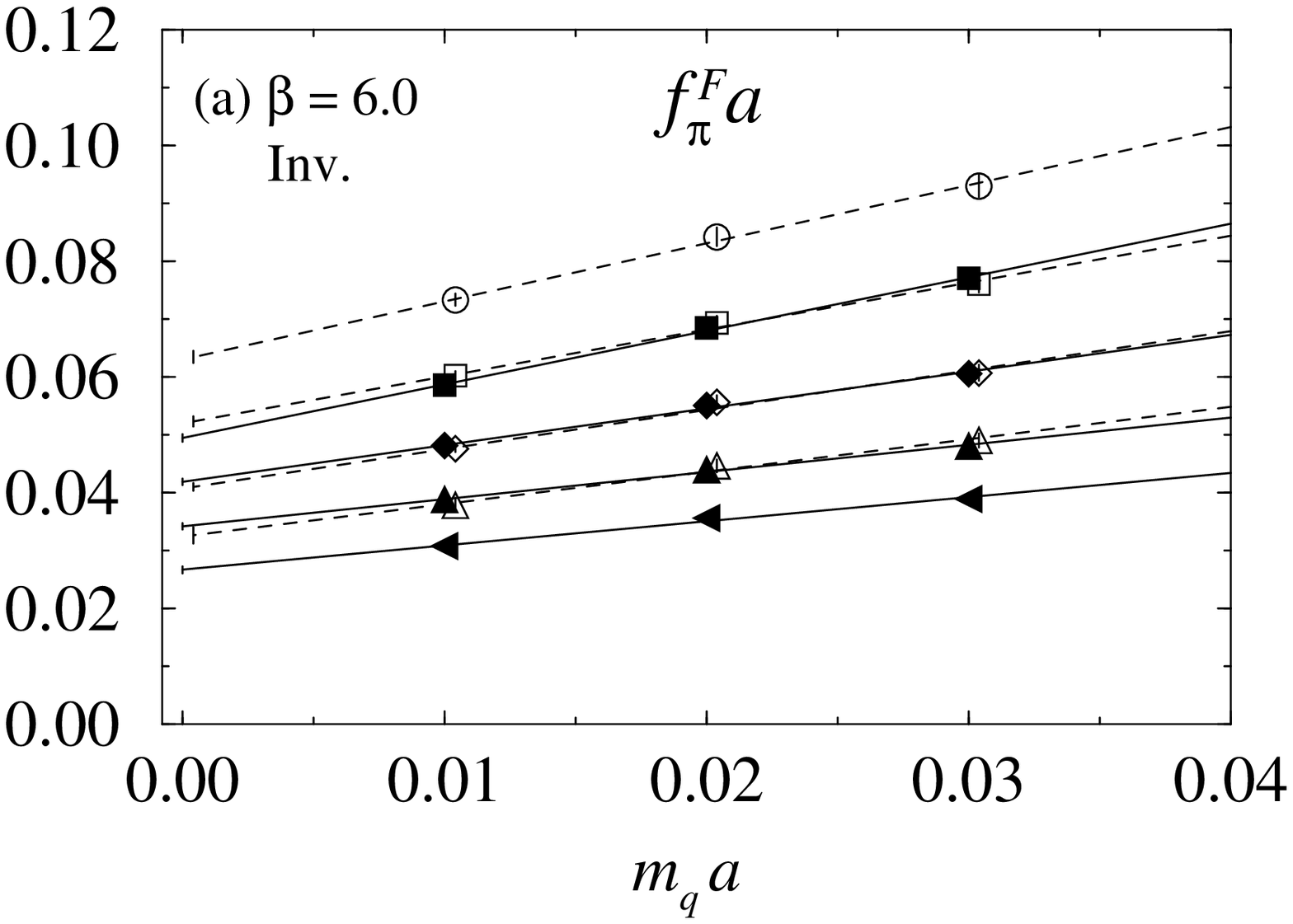} &
\epsfxsize=222pt\epsfbox{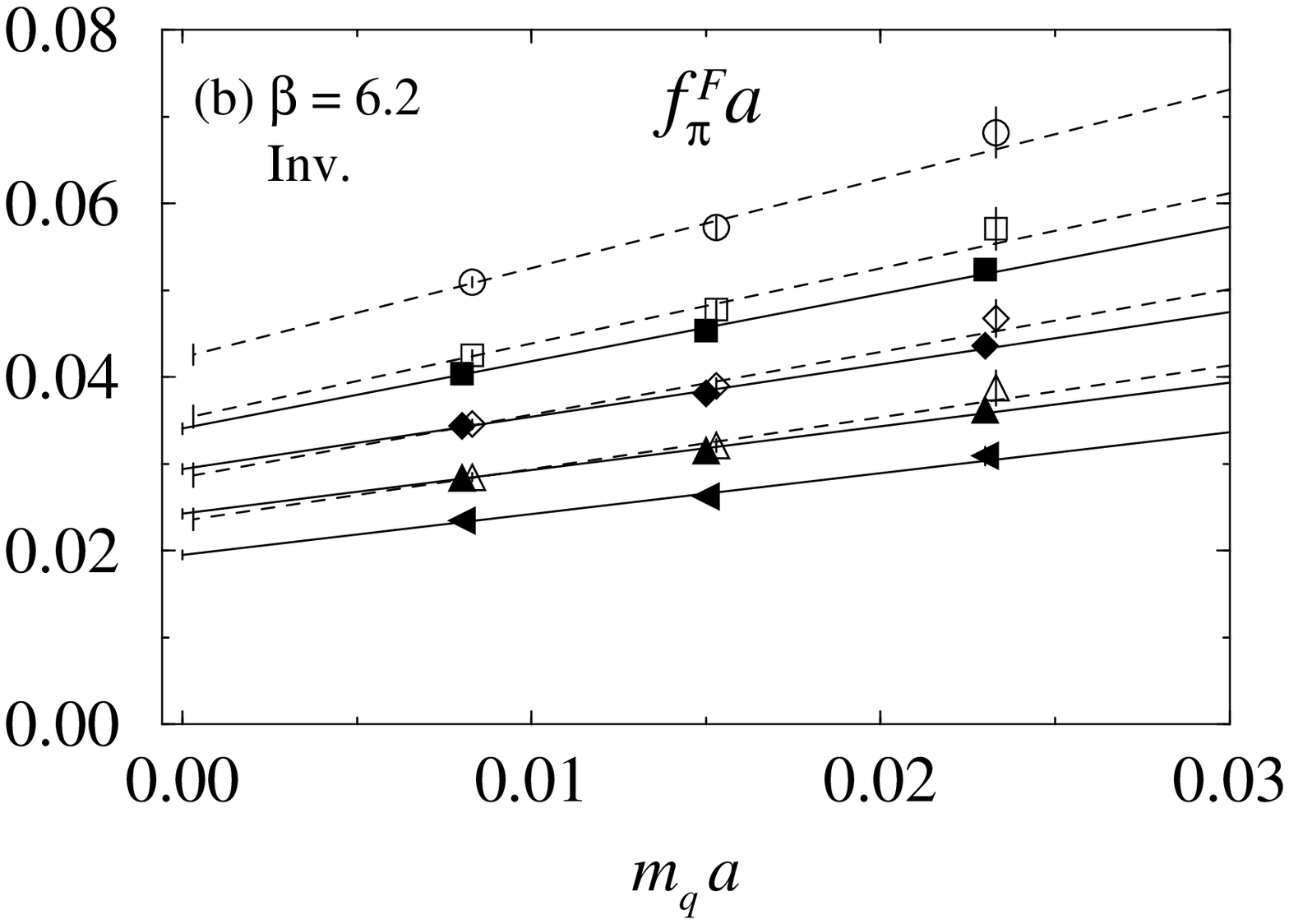} \\
\epsfxsize=222pt\epsfbox{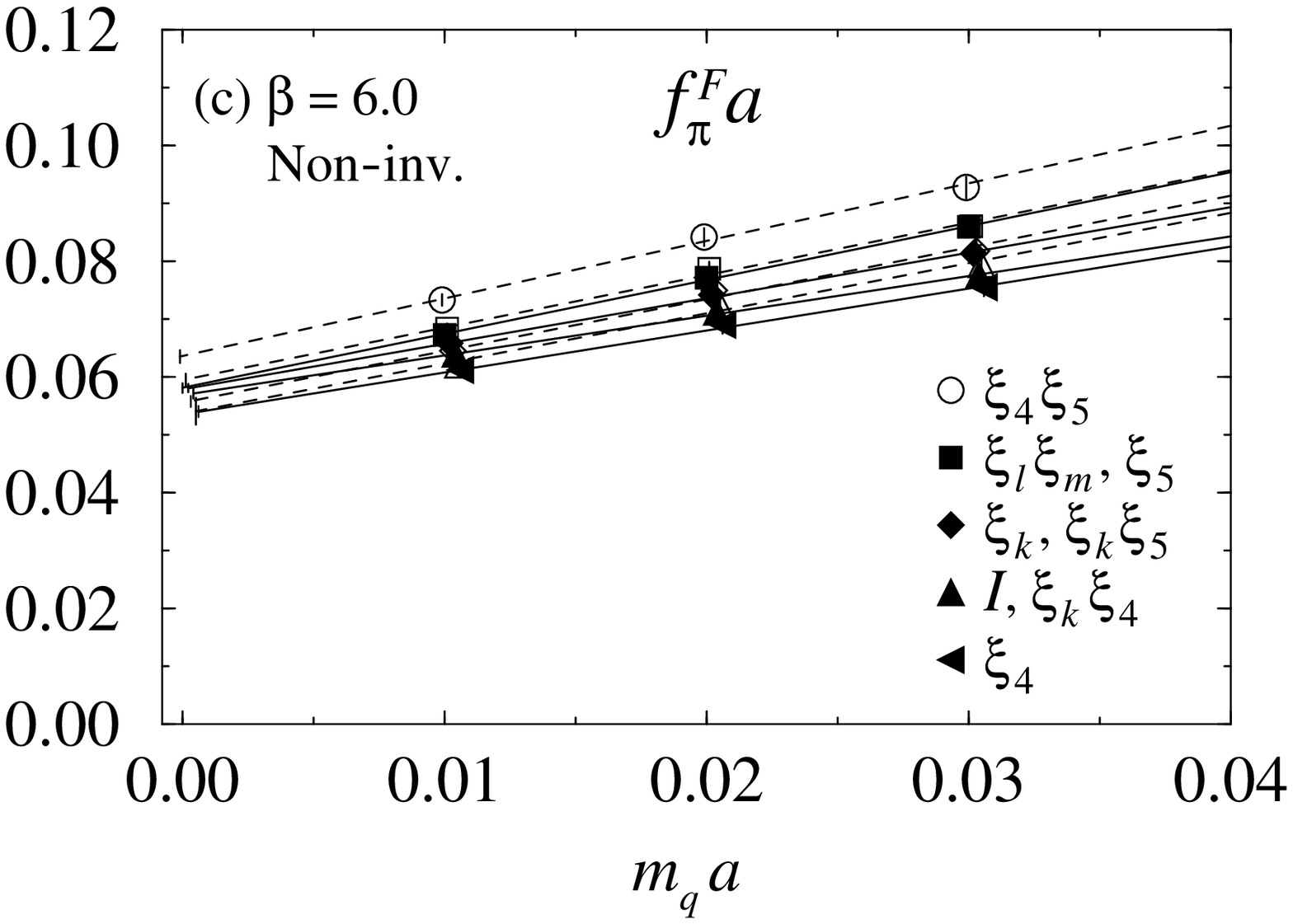} &
\epsfxsize=222pt\epsfbox{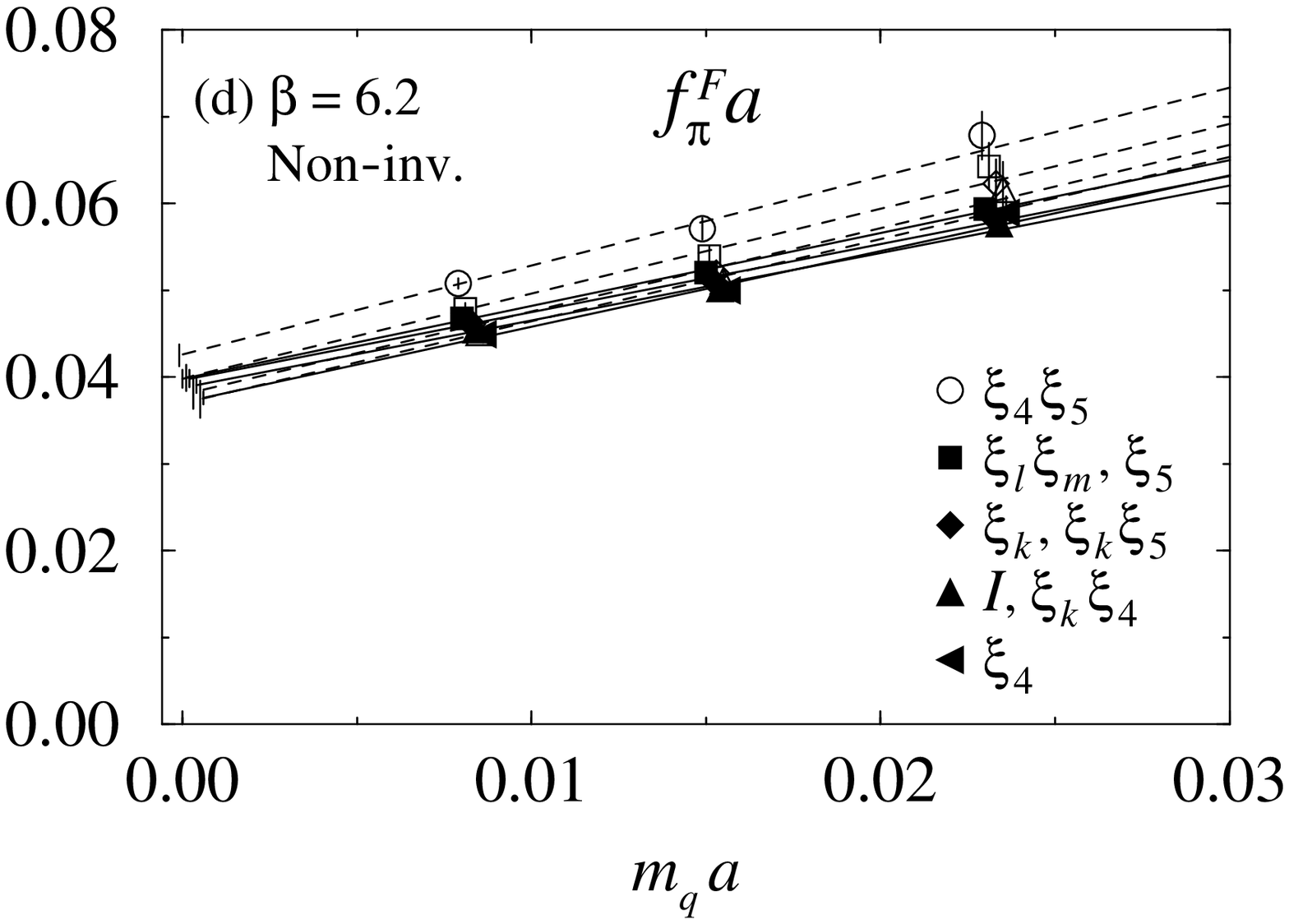} \\
\end{tabular}
\caption{Chiral behavior of the bare pion decay constants obtained by
         gauge invariant axial vector current for (a) $\beta = 6.0$
         and (b) $\beta = 6.2$, and by gauge non-invariant current
         for (c) $\beta = 6.0$ and (d) $\beta = 6.2$.  Omitted
         legends in the top two figures are same as that in the
         bottom figures.  Shape of symbols refer to the distance of
         the operator.  Some symbols denotes two flavors; the former
         one refers time-separated operators (filled symbols)
         including flavor $\xi_5$, and the latter one refers time-%
         local operators (open symbols).}
\label{fig:fpi1}
\end{figure}

In Fig.~\ref{fig:fpi1} we illustrate the chiral behavior of the bare
pion decay constants calculated with Eq.~(\ref{eq:fpi2}).  As with the
case for pion masses, we use a linear extrapolation toward the chiral
limit.

The pion decay constants obtained for eight irreducible representations
again form a degeneracy pattern, which, however, is different from that
for pion masses.  This is due to the fact that the pattern for the decay
constant reflects the distance of the axial vector current operator
rather than that of the pion operator: the two operators differ because
of the the Dirac factor, $\gamma_4\gamma_5$ for the axial vector current
and $\gamma_5$ for the pion.  We also observe that the KS flavor
dependence of the decay constant is much larger for the gauge invariant
operators than that for the non-invariant ones.  In contrast to the case
of mass, for which no renormalization is required and lattice symmetry
group controls, the pattern for pion decay constants mainly comes from
the insertion of gauge link variables, which is roughly written as
relation between the continuum and lattice axial vector currents:
\begin{equation}
A_\mu\big|_{\rm cont}\sim\left(\frac{1}{3}\langle{\rm Tr}\,U_\Box
\rangle\right)^{\!\mbox{$\frac{d-1}{4}$}}A_\mu^F\big|_{\rm lat}.
\end{equation}
Here $d$ is the distance of the axial vector current operator for the
gauge invariant case, while the non-invariant operator corresponds to
$d = 0$.

\begin{figure}[p]
\begin{tabular}{lc}
\epsfxsize=222pt\epsfbox{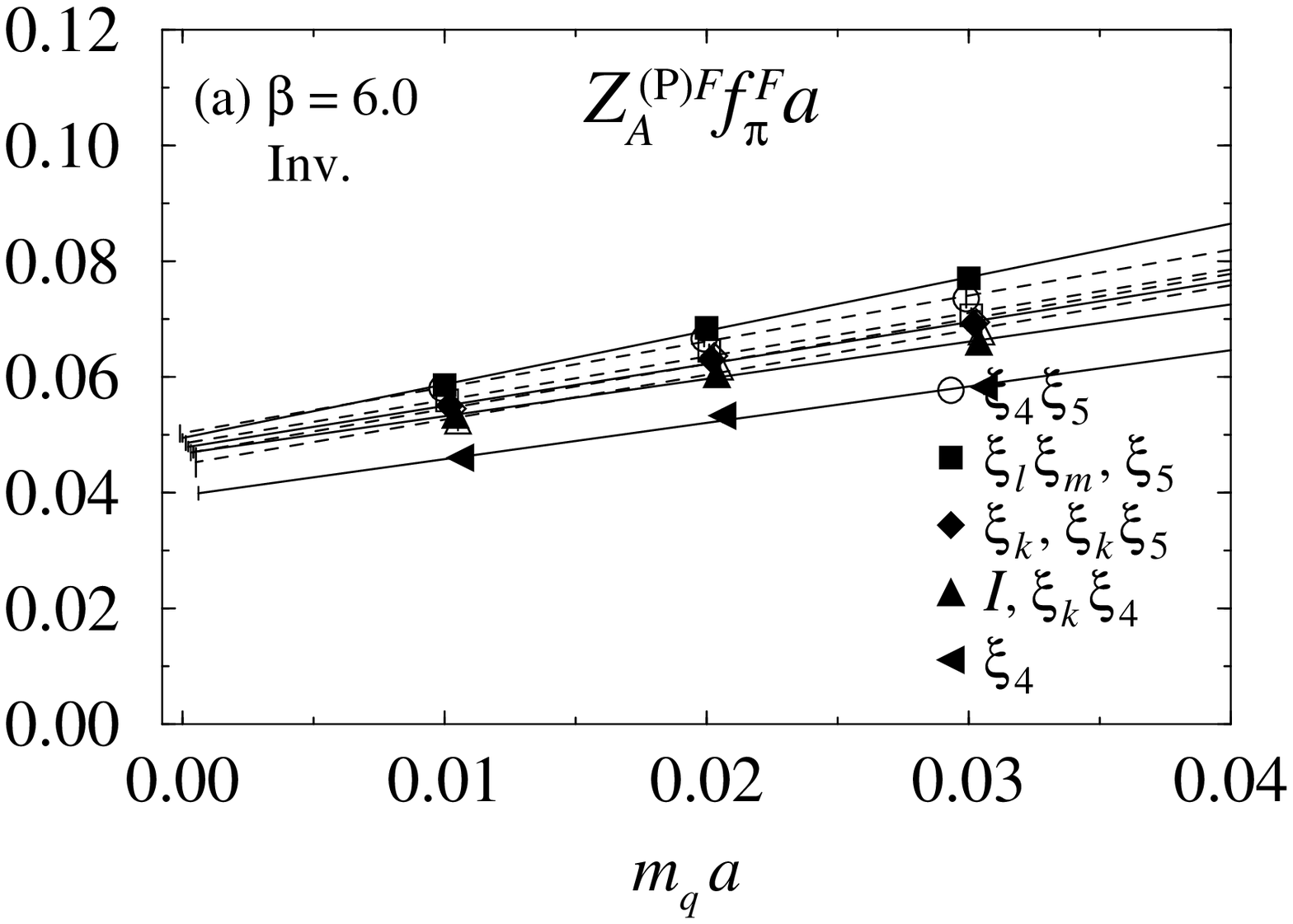} &
\epsfxsize=222pt\epsfbox{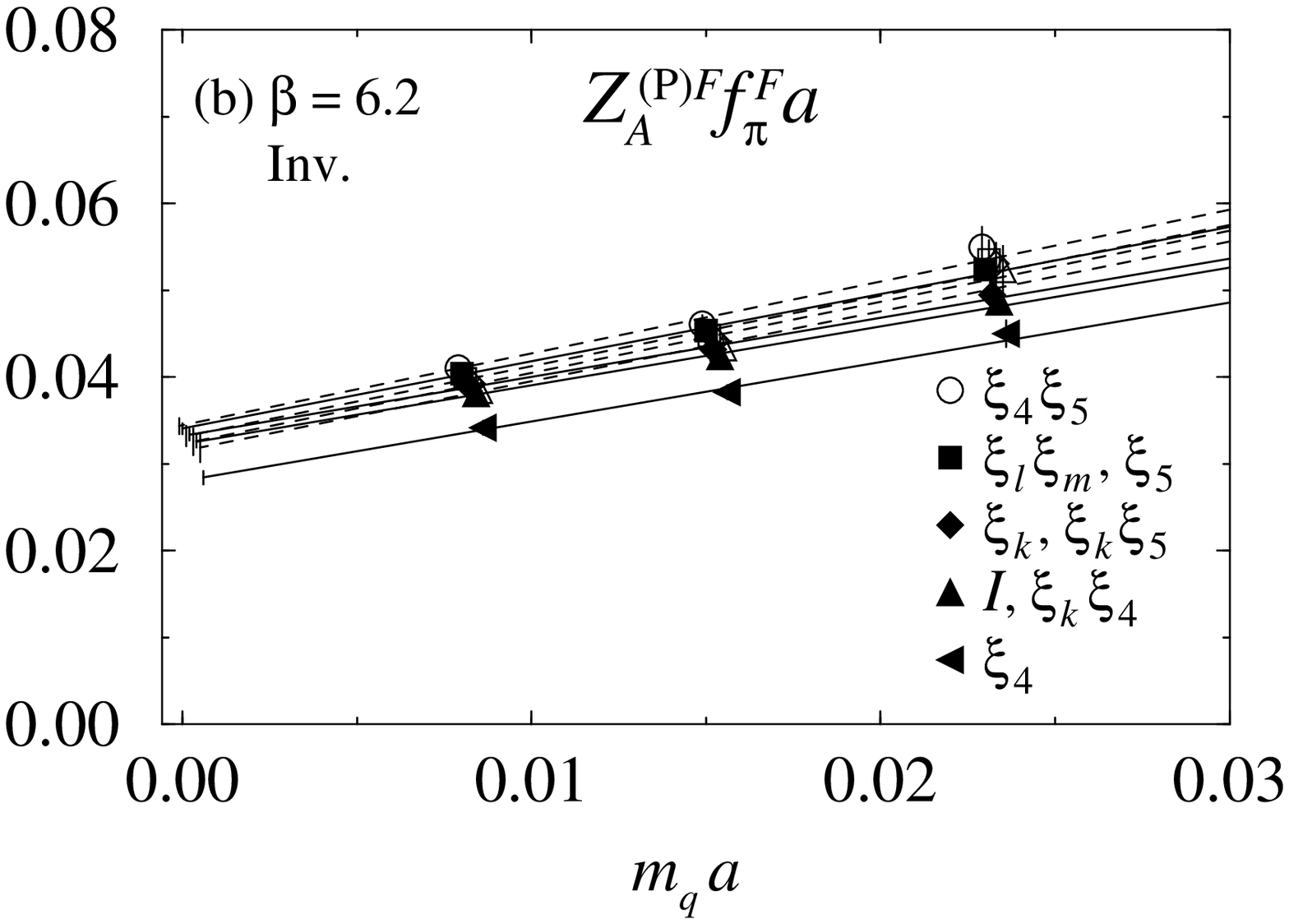} \\
\epsfxsize=222pt\epsfbox{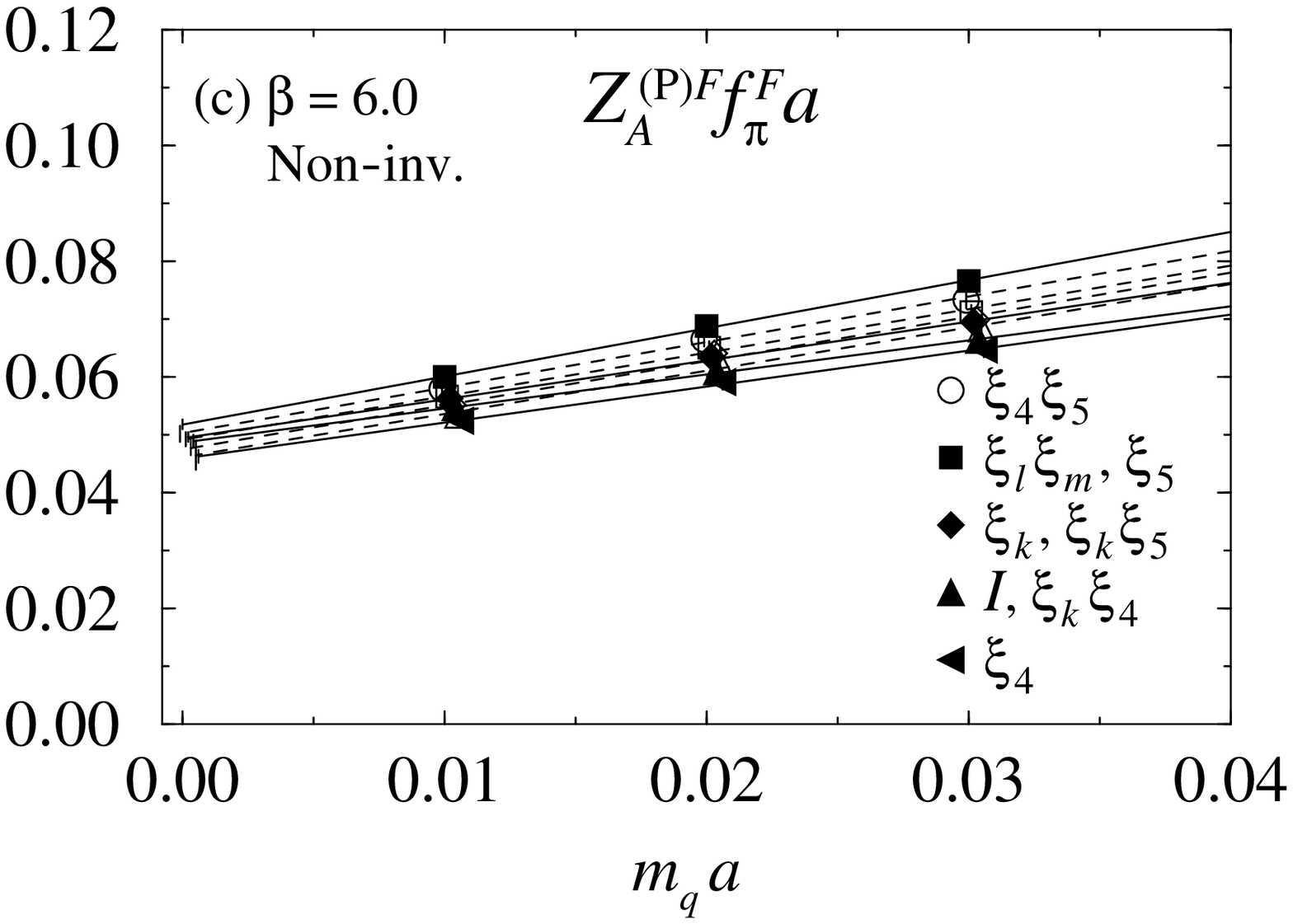} &
\epsfxsize=222pt\epsfbox{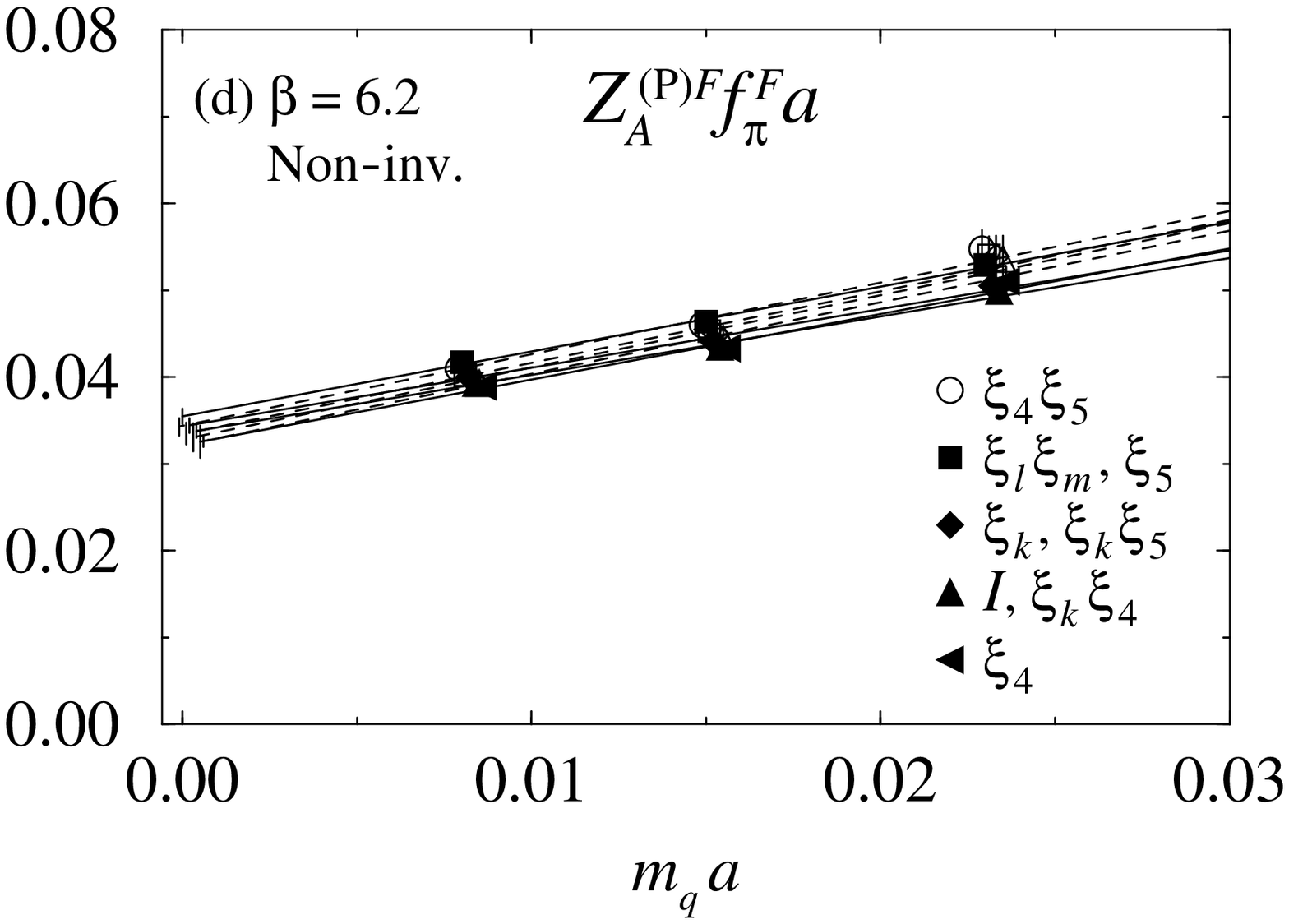} \\
\end{tabular}
\caption{Pion decay constant renormalized by one-loop perturbative 
         renormalization factor $Z_A^{(P)F}f_\pi^F$.  Figures (a)--%
         (d) correspond to those in Fig.~\ref{fig:fpi1}.}
\label{fig:fpi1P}
\end{figure}

\begin{figure}[p]
\begin{tabular}{lc}
\epsfxsize=222pt\epsfbox{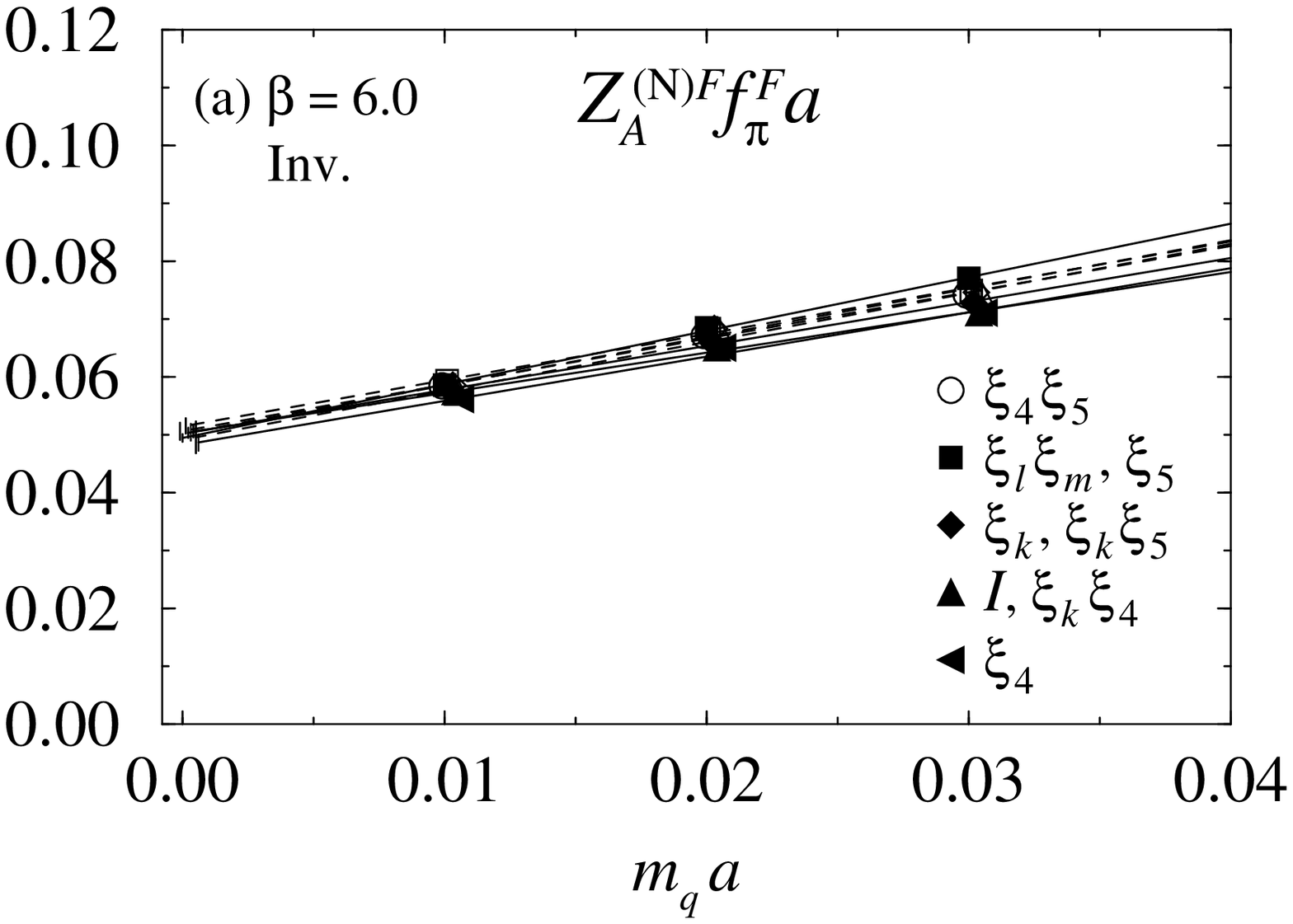} &
\epsfxsize=222pt\epsfbox{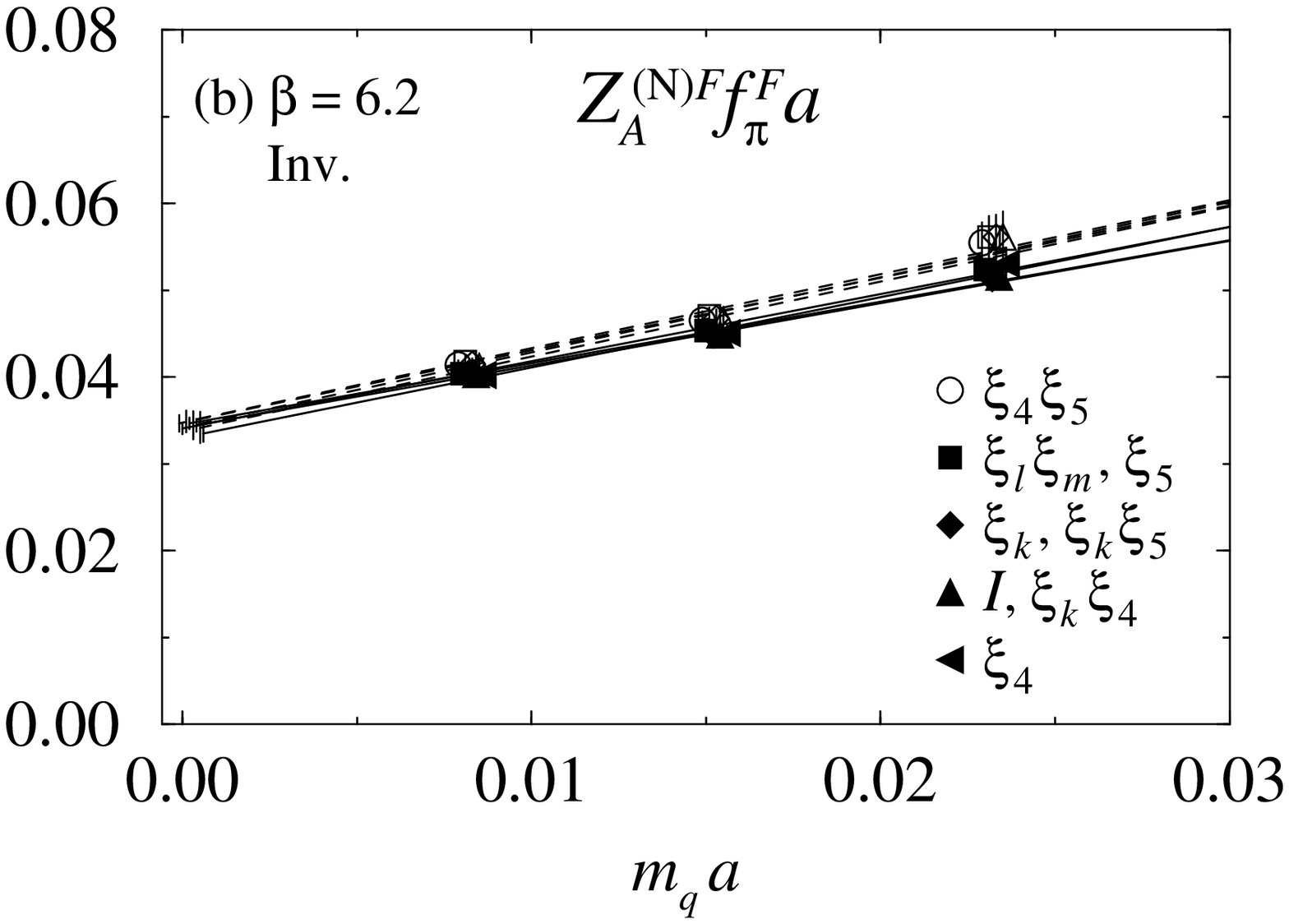} \\
\epsfxsize=222pt\epsfbox{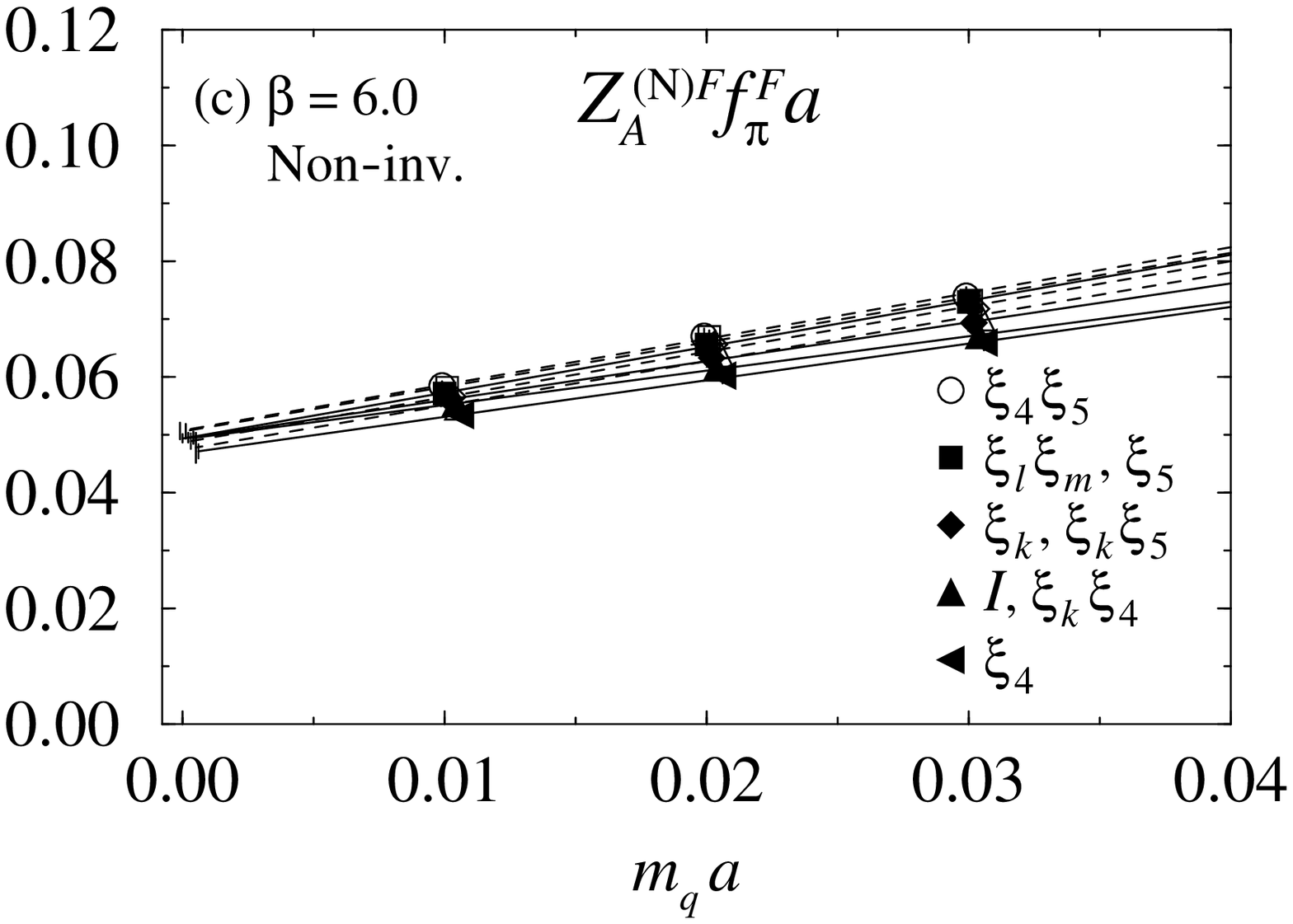} &
\epsfxsize=222pt\epsfbox{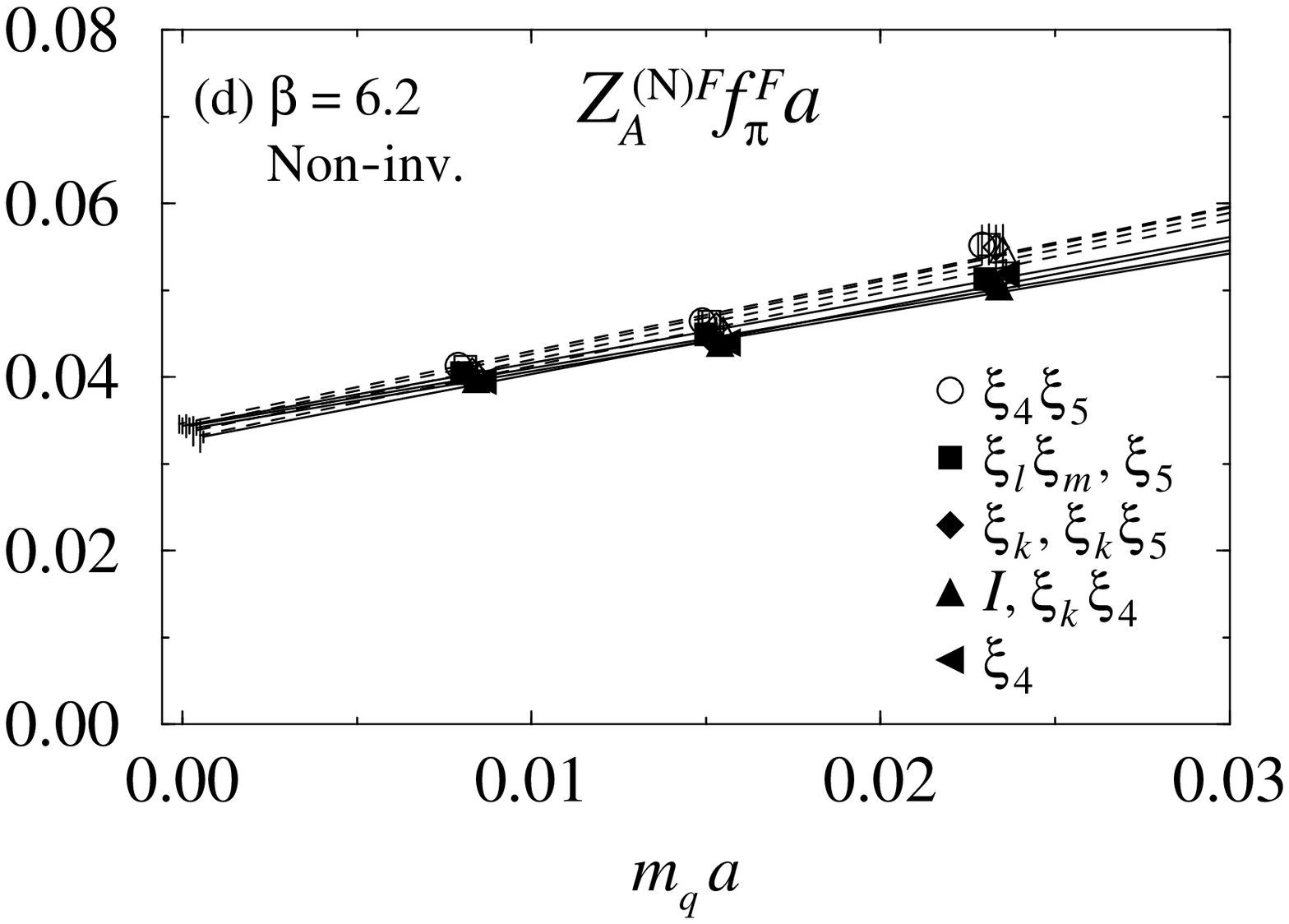} \\
\end{tabular}
\caption{Pion decay constant renormalized by non-perturbative
         renormalization factor $Z_A^{(N)F}f_\pi^F$.  Figures (a)--%
         (d) correspond to those in Fig.~\ref{fig:fpi1}.}
\label{fig:fpi1N}
\end{figure}

We show the decay constants after renormalization in Figs.~%
\ref{fig:fpi1P} and \ref{fig:fpi1N}.  With the use of perturbative
renormalization constants (Fig.~\ref{fig:fpi1P}), the discrepancy among
different KS flavor channels becomes smaller toward the continuum.  The
reduction of the discrepancy, however, is significantly more dramatic
with the use of non-perturbative renormalization constants as shown in
Fig.~\ref{fig:fpi1N}.  In particular, the large difference among bare
results obtained with gauge invariant operators almost disappears.

The numerical values for pion decay constants are collected in Tables~%
\ref{tab:fpi1}--\ref{tab:fpi1N}.  In contrast to the case of pion mass,
there is no flavor channel to give the same results for the gauge
invariant and non-invariant case, because the simultaneous local channel
does not exist for the axial vector current and the pion operator both
appearing in the calculation of the pion decay constant.

\section{Continuum extrapolation}
\label{sec:continuum}

\begin{figure}[p]
\epsfxsize=222pt\epsfbox{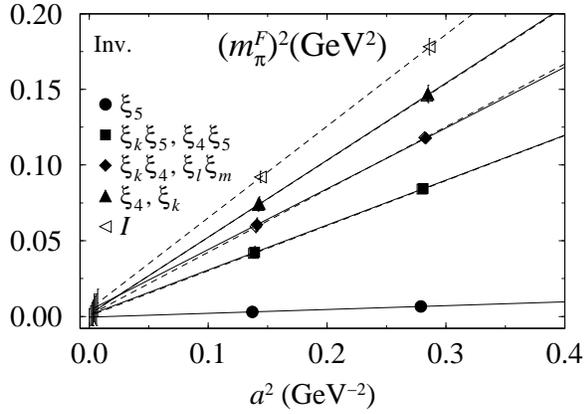}
\caption{Continuum limit of pion mass squared.  Symbols are same as
         those in Fig.~\ref{fig:mass1}.}
\label{fig:mass5}
\end{figure}

In Fig.~\ref{fig:mass5}, we present $a$-dependence of $(m_\pi^F)^2$
quadratically extrapolated to $m_q a = 0$, according to $O(a^2)$ scale
violation expected for the KS quark action.  We observe clear evidence
that the non-zero values of $(m_\pi^F)^2$ for the non-Nambu-Goldstone
channels vanish as $a^2$ toward the continuum limit, supporting the
restoration of full flavor symmetry of the KS action.

\begin{figure}[p]
\begin{tabular}{lc}
\epsfxsize=222pt\epsfbox{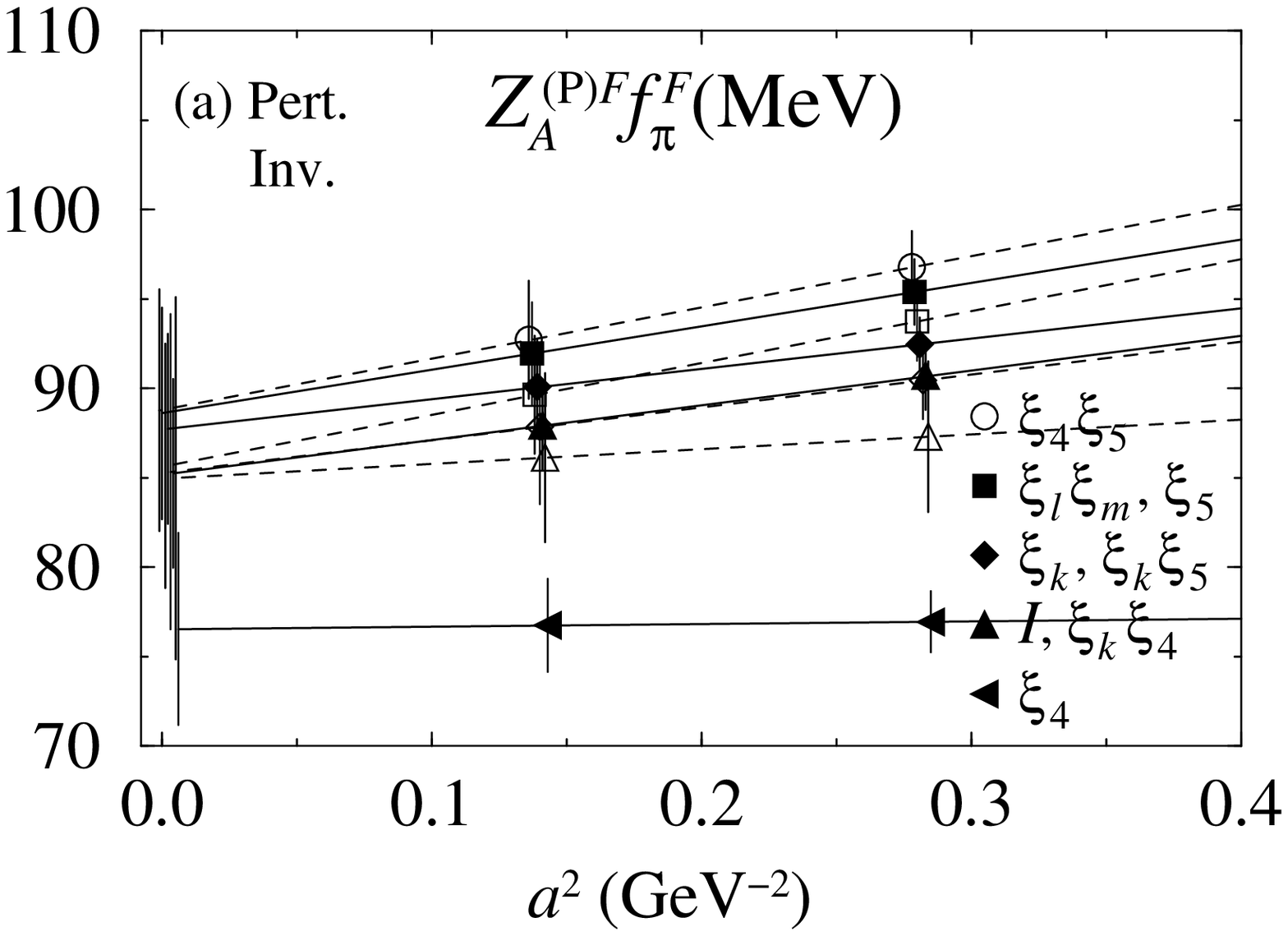} &
\epsfxsize=222pt\epsfbox{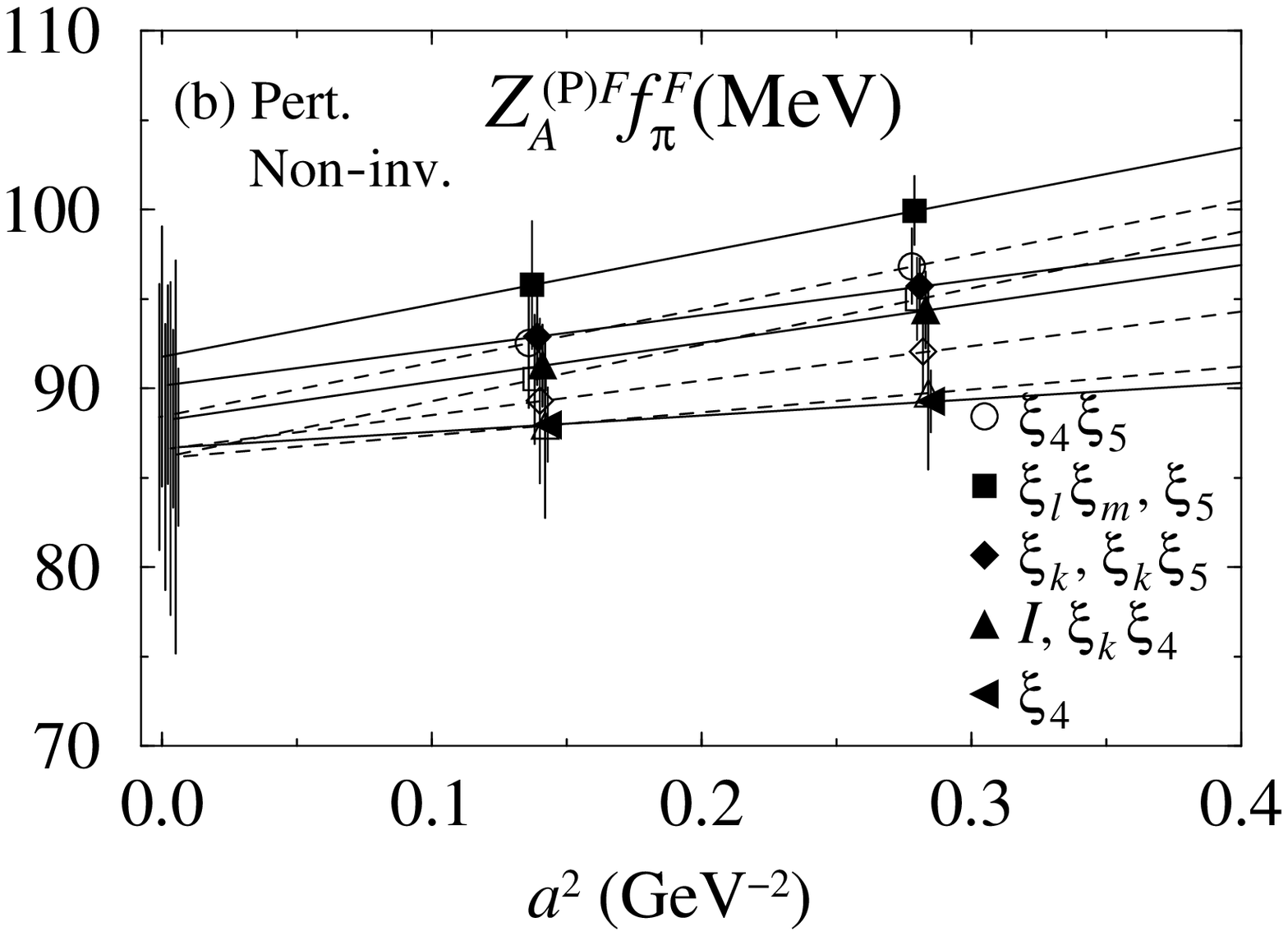} \\
\epsfxsize=222pt\epsfbox{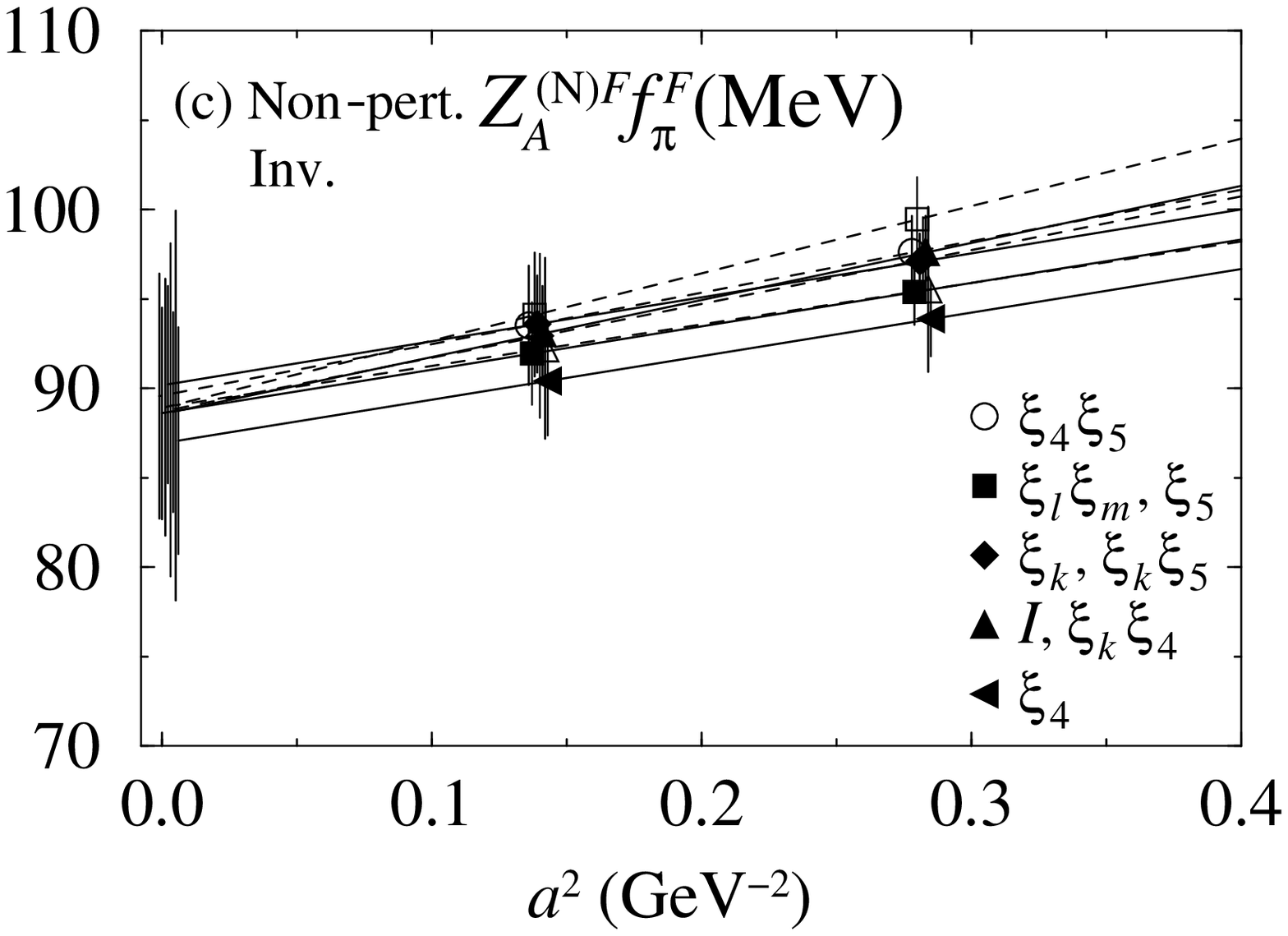} &
\epsfxsize=222pt\epsfbox{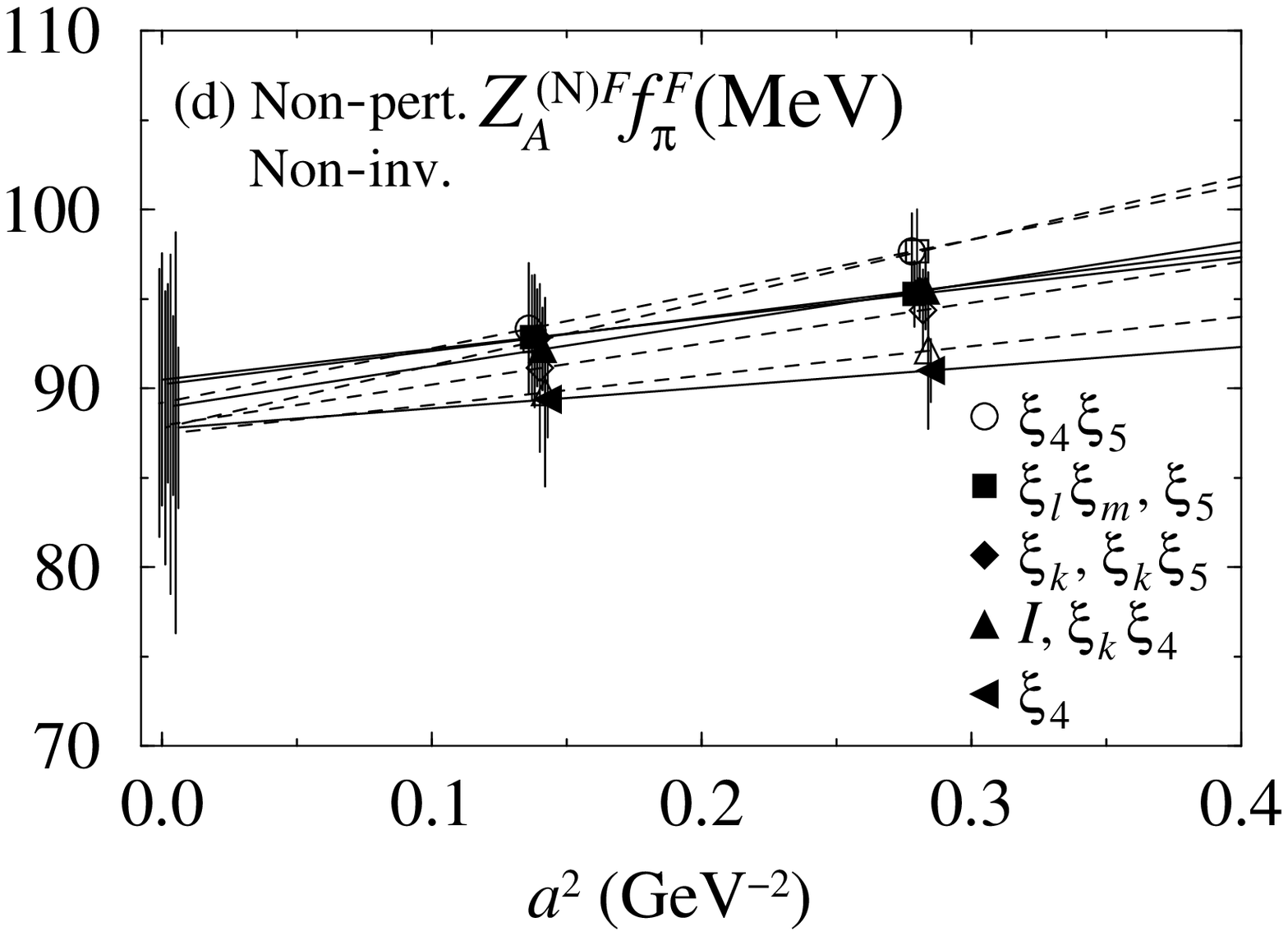} \\
\end{tabular}
\caption{Continuum limit of renormalized pion decay constants.
         Results obtained with perturbative renormalization factors 
         for (a) gauge-invariant and (b) non-invariant operator, and
         those with non-perturbative factors for (c) gauge-invariant
         and (d) non-invariant operator are shown.  Symbols are same
         as those in Figs.~\ref{fig:fpi1}--\ref{fig:fpi1N}.}
\label{fig:fpi5}
\end{figure}

The continuum extrapolation of the pion decay constant, renormalized
perturbatively or non-perturbatively, is shown in Fig.~\ref{fig:fpi5} as
a function of $a^2$.  In this figure with an enlarged vertical scale as
compared to Figs.~\ref{fig:fpi1P} and \ref{fig:fpi1N}, we observe a
general trend that the difference of values among various KS flavors
becomes smaller toward the continuum limit.  In particular, for non-%
perturbatively renormalized decay constants the central values in the
continuum limit agree within a 2\% accuracy, which is well below the
statistical errors of 5--10\%.  On the other hand, the convergence is
worse for the perturbatively renormalized decay constants.  The spread
in the continuum limit is 3--4\%, which is roughly the magnitude of
uncertainty one expects from higher-order corrections in the
renormalization factors. We consider that these results provide
evidence for both restoration of $\rm SU(4)_A$ flavor symmetry of the KS
action in the continuum limit and the effectiveness of the non-%
perturbatively evaluated renormalization constants.

The values of pion mass squared for various KS flavors are listed in
Table~\ref{tab:mass5}, and those for pion decay constants are collected
in Tables~\ref{tab:fpi5P} and \ref{tab:fpi5N}. As our best value for the
decay constant, we take $f_\pi = 89(6)$ MeV obtained with the gauge
invariant axial vector current in the $\xi_5$ channel which requires no
renormalization.  This value is compared with the experiment 92.4(3) MeV%
\cite{PDG_98}.  Possible quenching errors are not visible within the
statistical error of 6 MeV. 

Let us recall that the decay constant in the $\xi_5$ channel can also be
calculated from the pion operator using Eqs.~(\ref{eq:fpiP2}) and
(\ref{eq:fpi3}).  Results are added in the bottom lines of Table~%
\ref{tab:fpi5P} (and \ref{tab:fpi5N} for convenience of readers), which
show reasonable agreement with those from the axial vector current in
the $\xi_5$ channel, as expected. 

\section{Conclusion}
\label{sec:conclusion}

In this article we have presented an analysis of the pion decay constant
in quenched QCD using the Kogut-Susskind quark action.  Our best
estimate for the decay constant in the continuum limit is 89(6) MeV,
which is obtained with the gauge invariant axial vector current which
respects $\rm U(1)_A$ symmetry.

We have carried out a detailed comparison of perturbative and non-%
perturbative axial vector renormalization treatments.  We conclude that
the non-perturbative renormalization factors efficiently eliminate the
flavor breaking effect in the decay constant in the continuum limit,
while an apparent flavor-dependent difference still remains with the
perturbative factors.

\section*{Acknowledgments}

This work is supported by the Supercomputer Project No.45 (FY1999) of
High Energy Accelerator Research Organization (KEK), and also in part by
the Grants-in-Aid of the Ministry of Education (Nos. 09304029, 10640246,
10640248, 10740107, 10740125, 11640294, 11740162).  K-I.I. is supported
by the JSPS Research Fellowship.

\widetext

\pagebreak

\pagebreak

\narrowtext

\widetext

\newpage

\narrowtext

\begin{table}
\caption{Calculation parameters for evaluation of non-perturbative
         renormalization constants.}
\label{tab:parameter}
\begin{tabular}{lcccc}
\multicolumn{1}{c}{$\beta$}
    & \multicolumn{1}{c}{$L^3\times T$}
    & \multicolumn{1}{c}{$m_q a$}
    & \multicolumn{1}{c}{$a^{-1}$(GeV)}
    & \multicolumn{1}{c}{\#conf.}                          \\
\hline
6.0 & $32^3\times 32$ & 0.010, 0.020, 0.030 & 1.88(4) & 30 \\
6.2 & $32^3\times 32$ & 0.008, 0.015, 0.023 & 2.65(9) & 30 \\
\end{tabular}
\end{table}

\begin{table}
\caption{Renormalization constants $Z_A^F$ used for renormalizing
         pion decay constants.}
\label{tab:Zs}

\begin{tabular}{lllll}
\multicolumn{5}{c}{(a) $\beta=6.0$}                                 \\
\hline
                          & \multicolumn{2}{c}{Perturbative}
                          & \multicolumn{2}{c}{Non-perturbative}    \\
\multicolumn{1}{c}{Operator}
                          & \multicolumn{1}{c}{Gauge inv.}
                          & \multicolumn{1}{c}{Non-inv.}
                          & \multicolumn{1}{c}{Gauge inv.}
                          & \multicolumn{1}{c}{Non-inv.}            \\
\hline
$(\ggox {4}{5}{5}      )$ & 1      & 0.8917 & 1         & 0.85019(7)\\
$(\ggoxx{4}{5}{k}   {5})$ & 1.1436 & 0.8547 & 1.2008(1) & 0.8527(1) \\
$(\ggoxx{4}{5}{k}   {4})$ & 1.3749 & 0.8556 & 1.4799(2) & 0.8656(1) \\
$(\ggox {4}{5}{4}      )$ & 1.4950 & 0.8569 & 1.8242(3) & 0.8736(2) \\
$(\ggoxx{4}{5}{4}   {5})$ & 0.7908 & 0.7908 & 0.7976(2) & 0.7976(2) \\
$(\ggoxx{4}{5}{\ell}{m})$ & 0.9294 & 0.8277 & 0.9860(1) & 0.8508(1) \\
$(\ggox {4}{5}{k}      )$ & 1.1440 & 0.8550 & 1.2294(3) & 0.8767(2) \\
$(\ggoI {4}{5}         )$ & 1.3837 & 0.8605 & 1.5145(5) & 0.8835(2) \\
\end{tabular}

\begin{tabular}{lllll}
\multicolumn{5}{c}{(b) $\beta=6.2$}                                 \\
\hline
                          & \multicolumn{2}{c}{Perturbative}
                          & \multicolumn{2}{c}{Non-perturbative}    \\
\multicolumn{1}{c}{Operator}
                          & \multicolumn{1}{c}{Gauge inv.}
                          & \multicolumn{1}{c}{Non-inv.}
                          & \multicolumn{1}{c}{Gauge inv.}
                          & \multicolumn{1}{c}{Non-inv.}            \\
\hline
$(\ggox {4}{5}{5}      )$ & 1      & 0.8917 & 1         & 0.86430(7)\\
$(\ggoxx{4}{5}{k}   {5})$ & 1.1338 & 0.8643 & 1.1783(1) & 0.86363(7)\\
$(\ggoxx{4}{5}{k}   {4})$ & 1.3434 & 0.8651 & 1.4221(2) & 0.8739(2) \\
$(\ggox {4}{5}{4}      )$ & 1.4567 & 0.8663 & 1.7164(3) & 0.8803(2) \\
$(\ggoxx{4}{5}{4}   {5})$ & 0.8065 & 0.8065 & 0.8136(1) & 0.8136(1) \\
$(\ggoxx{4}{5}{\ell}{m})$ & 0.9369 & 0.8401 & 0.9838(1) & 0.8600(1) \\
$(\ggox {4}{5}{k}      )$ & 1.1342 & 0.8646 & 1.1999(1) & 0.8825(2) \\
$(\ggoI {4}{5}         )$ & 1.3508 & 0.8696 & 1.4472(4) & 0.8882(2) \\
\end{tabular}
\end{table}

\begin{table}
\caption{Calculation parameters of our simulation.}
\label{tab:parameter2}
\begin{tabular}{lcccc}
\multicolumn{1}{c}{$\beta$}
    & \multicolumn{1}{c}{$L^3\times T$}
    & \multicolumn{1}{c}{$m_q a$}
    & \multicolumn{1}{c}{$a^{-1}$(GeV)}
    & \multicolumn{1}{c}{\#conf.}                          \\
\hline
6.0 & $32^3\times 64$ & 0.010, 0.020, 0.030 & 1.92(2) & 100\\
6.2 & $48^3\times 64$ & 0.008, 0.015, 0.023 & 2.70(5) & 60 \\
\end{tabular}
\end{table}

\mediumtext

\begin{table}[p]
\caption{Minimum time slice $t_{\rm min}$ common for all flavors
         except for $\xi_F = \xi_4$ in the parenthesis (See text for reason),
         and $\chi^2 / N_{\rm DF}$ of global fits for the local
         channel.}
\label{tab:range}
\begin{tabular}{lccccccccc}
    &
    & \multicolumn{2}{c}{$\langle A_4^F(t)\pi_W^F(0)\rangle$}
    & \multicolumn{2}{c}{$\langle\pi^F(t)\pi_W^F(0)\rangle$}
    & \multicolumn{2}{c}{$\langle\pi_W^F(t)\pi_W^F(0)\rangle$}
    & \multicolumn{2}{c}{$\langle\rho_k^F(t)\rho_{kW}^F(0)\rangle$}\\
\multicolumn{1}{c}{$\beta$}
    & \multicolumn{1}{c}{$m_q a$}
    & \multicolumn{1}{c}{$t_{\rm min}$}
    & \multicolumn{1}{c}{$\chi^2/N_{\rm DF}$}
    & \multicolumn{1}{c}{$t_{\rm min}$}
    & \multicolumn{1}{c}{$\chi^2/N_{\rm DF}$}
    & \multicolumn{1}{c}{$t_{\rm min}$}
    & \multicolumn{1}{c}{$\chi^2/N_{\rm DF}$}
    & \multicolumn{1}{c}{$t_{\rm min}$}
    & \multicolumn{1}{c}{$\chi^2/N_{\rm DF}$}                   \\
\hline
6.0 & 0.030 & 17 & 1.37 & 17 & 1.24 & 14     & 1.07 & 18 & 1.34 \\
    & 0.020 & 17 & 1.05 & 17 & 0.95 & 15     & 1.27 & 17 & 0.87 \\
    & 0.010 & 16 & 0.97 & 16 & 0.99 & 15     & 0.83 & 15 & 1.27 \\
6.2 & 0.023 & 17 & 1.40 & 24 & 1.06 & 17(16) & 0.92 & 23 & 0.79 \\
    & 0.015 & 16 & 1.07 & 23 & 0.85 & 19(19) & 0.97 & 23 & 0.94 \\
    & 0.008 & 15 & 1.33 & 22 & 0.99 & 19(20) & 1.21 & 22 & 0.58 \\
\end{tabular}
\end{table}

\widetext

\begin{table}[p]
\squeezetable
\caption{Pion mass squared $(m_\pi^F)^2$ in lattice units.  Note that the
        correlation function with the local pion operator in the $\xi_5$
        channel gives exactly the same results
        for the gauge invariant and non-invariant case.}
\label{tab:mass1}

\begin{tabular}{lllllllll}
\multicolumn{9}{c}{(a) $\beta=6.0$}                                                                           \\
\hline
                      & \multicolumn{4}{c}{Gauge invariant}
                      & \multicolumn{4}{c}{Non-invariant}                                                     \\
\multicolumn{1}{c}{Operator}
                      & \multicolumn{1}{c}{$m_q a=0.030$}
                      & \multicolumn{1}{c}{$m_q a=0.020$}
                      & \multicolumn{1}{c}{$m_q a=0.010$}
                      & \multicolumn{1}{c}{$m_q a\rightarrow 0$}
                      & \multicolumn{1}{c}{$m_q a=0.030$}
                      & \multicolumn{1}{c}{$m_q a=0.020$}
                      & \multicolumn{1}{c}{$m_q a=0.010$}
                      & \multicolumn{1}{c}{$m_q a\rightarrow 0$}                                              \\
\hline
$(\gox {5}{5}      )$ & 0.1687(3)& 0.1129(3)& 0.0575(2)& 0.0018(2)& \ \ $\longleftarrow$
                                                                  & \ \ $\longleftarrow$
                                                                  & \ \ $\longleftarrow$
                                                                  & \ \ $\longleftarrow$                      \\
$(\goxx{5}{k}   {5})$ & 0.2077(4)& 0.1454(4)& 0.0846(4)& 0.0228(4)& 0.2077(4)& 0.1454(4)& 0.0846(4)& 0.0228(4)\\
$(\goxx{5}{k}   {4})$ & 0.2194(4)& 0.1561(5)& 0.0946(6)& 0.0317(6)& 0.2194(4)& 0.1562(5)& 0.0947(6)& 0.0317(5)\\
$(\gox {5}{4}      )$ & 0.2260(5)& 0.1630(6)& 0.1023(8)& 0.0396(9)& 0.2261(5)& 0.1630(6)& 0.1024(8)& 0.0396(9)\\
$(\goxx{5}{4}   {5})$ & 0.2086(5)& 0.1459(5)& 0.0848(5)& 0.0226(5)& 0.2087(4)& 0.1460(5)& 0.0848(5)& 0.0226(5)\\
$(\goxx{5}{\ell}{m})$ & 0.2203(6)& 0.1567(6)& 0.0948(5)& 0.0318(4)& 0.2203(5)& 0.1567(5)& 0.0947(5)& 0.0317(3)\\
$(\gox {5}{k}      )$ & 0.2268(6)& 0.1633(7)& 0.1021(7)& 0.0393(4)& 0.2268(6)& 0.1634(6)& 0.1021(7)& 0.0392(4)\\
$(\goI {5}         )$ & 0.2324(8)& 0.170(1) & 0.110(1) & 0.048(1) & 0.2325(7)& 0.1699(9)& 0.110(1) & 0.048(1) \\
\end{tabular}

\begin{tabular}{lllllllll}
\multicolumn{9}{c}{(b) $\beta=6.2$}                                                                           \\
\hline
                      & \multicolumn{4}{c}{Gauge invariant}
                      & \multicolumn{4}{c}{Non-invariant}                                                     \\
\multicolumn{1}{c}{Operator}
                      & \multicolumn{1}{c}{$m_q a=0.023$}
                      & \multicolumn{1}{c}{$m_q a=0.015$}
                      & \multicolumn{1}{c}{$m_q a=0.008$}
                      & \multicolumn{1}{c}{$m_q a\rightarrow 0$}
                      & \multicolumn{1}{c}{$m_q a=0.023$}
                      & \multicolumn{1}{c}{$m_q a=0.015$}
                      & \multicolumn{1}{c}{$m_q a=0.008$}
                      & \multicolumn{1}{c}{$m_q a\rightarrow 0$}                                             \\
\hline
$(\gox {5}{5}      )$ & 0.0927(3)& 0.0604(2)& 0.0326(3)& 0.0004(4)& \ \ $\longleftarrow$
                                                                  & \ \ $\longleftarrow$
                                                                  & \ \ $\longleftarrow$
                                                                  & \ \ $\longleftarrow$                      \\
$(\goxx{5}{k}   {5})$ & 0.1017(3)& 0.0679(3)& 0.0394(3)& 0.0058(4)& 0.1017(3)& 0.0679(3)& 0.0393(3)& 0.0058(4)\\
$(\goxx{5}{k}   {4})$ & 0.1046(4)& 0.0706(3)& 0.0420(4)& 0.0083(4)& 0.1046(4)& 0.0706(3)& 0.0420(4)& 0.0083(4)\\
$(\gox {5}{4}      )$ & 0.1062(4)& 0.0724(3)& 0.0438(4)& 0.0102(4)& 0.1063(4)& 0.0723(3)& 0.0438(4)& 0.0102(4)\\
$(\goxx{5}{4}   {5})$ & 0.1019(3)& 0.0680(3)& 0.0393(3)& 0.0057(4)& 0.1021(4)& 0.0681(3)& 0.0394(3)& 0.0056(3)\\
$(\goxx{5}{\ell}{m})$ & 0.1047(3)& 0.0706(3)& 0.0419(3)& 0.0081(4)& 0.1049(4)& 0.0706(3)& 0.0420(4)& 0.0081(4)\\
$(\gox {5}{k}      )$ & 0.1064(4)& 0.0724(3)& 0.0438(3)& 0.0101(4)& 0.1066(4)& 0.0724(3)& 0.0439(4)& 0.0100(4)\\
$(\goI {5}         )$ & 0.1080(4)& 0.0742(3)& 0.0461(4)& 0.0127(4)& 0.1082(4)& 0.0742(4)& 0.0461(5)& 0.0125(4)\\
\end{tabular}
\end{table}

\begin{table}[p]
\squeezetable
\caption{Bare pion decay constant $f_\pi^F$ in lattice units.
         The bottom line shows results obtained from pion operator
         $f_\pi^{(P)5}$.}
\label{tab:fpi1}

\begin{tabular}{lllllllll}
\multicolumn{9}{c}{(a) $\beta=6.0$}                                                                               \\
\hline
                          & \multicolumn{4}{c}{Gauge invariant}
                          & \multicolumn{4}{c}{Non-invariant}                                                     \\
\multicolumn{1}{c}{Operator}
                          & \multicolumn{1}{c}{$m_q a=0.030$}
                          & \multicolumn{1}{c}{$m_q a=0.020$}
                          & \multicolumn{1}{c}{$m_q a=0.010$}
                          & \multicolumn{1}{c}{$m_q a\rightarrow 0$}
                          & \multicolumn{1}{c}{$m_q a=0.030$}
                          & \multicolumn{1}{c}{$m_q a=0.020$}
                          & \multicolumn{1}{c}{$m_q a=0.010$}
                          & \multicolumn{1}{c}{$m_q a\rightarrow 0$}                                              \\
\hline
$(\ggox {4}{5}{5}      )$ & 0.0770(5)& 0.0686(6)& 0.0586(4)& 0.0495(6)& 0.0859(6)& 0.0772(6)& 0.0673(6)& 0.0582(8)\\
$(\ggoxx{4}{5}{k}   {5})$ & 0.0606(8)& 0.0551(6)& 0.0482(4)& 0.0420(5)& 0.081(1) & 0.0743(9)& 0.0659(6)& 0.0581(7)\\
$(\ggoxx{4}{5}{k}   {4})$ & 0.0481(8)& 0.0440(7)& 0.0389(4)& 0.0342(5)& 0.078(1) & 0.071(1) & 0.0640(7)& 0.0572(9)\\
$(\ggox {4}{5}{4}      )$ & 0.0390(8)& 0.0357(8)& 0.0308(4)& 0.0267(6)& 0.076(2) & 0.069(2) & 0.0613(7)& 0.0541(9)\\
$(\ggoxx{4}{5}{4}   {5})$ & 0.093(2) & 0.084(2) & 0.073(1) & 0.064(1) & 0.093(2) & 0.084(2) & 0.073(1) & 0.064(1) \\
$(\ggoxx{4}{5}{\ell}{m})$ & 0.076(1) & 0.069(1) & 0.0603(8)& 0.0524(9)& 0.086(1) & 0.079(1) & 0.0684(8)& 0.060(1) \\
$(\ggox {4}{5}{k}      )$ & 0.061(1) & 0.056(1) & 0.0477(6)& 0.0410(7)& 0.082(2) & 0.075(1) & 0.0646(8)& 0.0559(9)\\
$(\ggoI {4}{5}         )$ & 0.049(1) & 0.045(1) & 0.038(1) & 0.033(1) & 0.080(2) & 0.072(2) & 0.062(2) & 0.054(2) \\
\hline
$(\gox  {5}   {5}      )$ & 0.0789(6)& 0.0697(6)& 0.0602(5)& 0.0509(6)& \ \ $\longleftarrow$
                                                                      & \ \ $\longleftarrow$
                                                                      & \ \ $\longleftarrow$
                                                                      & \ \ $\longleftarrow$                      \\
\end{tabular}

\begin{tabular}{lllllllll}
\multicolumn{9}{c}{(b) $\beta=6.2$}                                                                               \\
\hline
                          & \multicolumn{4}{c}{Gauge invariant}
                          & \multicolumn{4}{c}{Non-invariant}                                                     \\
\multicolumn{1}{c}{Operator}
                          & \multicolumn{1}{c}{$m_q a=0.023$}
                          & \multicolumn{1}{c}{$m_q a=0.015$}
                          & \multicolumn{1}{c}{$m_q a=0.008$}
                          & \multicolumn{1}{c}{$m_q a\rightarrow 0$}
                          & \multicolumn{1}{c}{$m_q a=0.023$}
                          & \multicolumn{1}{c}{$m_q a=0.015$}
                          & \multicolumn{1}{c}{$m_q a=0.008$}
                          & \multicolumn{1}{c}{$m_q a\rightarrow 0$}                                              \\
\hline
$(\ggox {4}{5}{5}      )$ & 0.0524(8)& 0.0454(4)& 0.0404(4)& 0.0341(6)& 0.0594(6)& 0.0520(5)& 0.0468(6)& 0.040(1) \\
$(\ggoxx{4}{5}{k}   {5})$ & 0.044(1) & 0.0381(6)& 0.0344(5)& 0.0294(6)& 0.058(1) & 0.0511(7)& 0.0463(6)& 0.040(1) \\
$(\ggoxx{4}{5}{k}   {4})$ & 0.0362(9)& 0.0315(6)& 0.0284(4)& 0.0243(5)& 0.058(1) & 0.0502(8)& 0.0455(6)& 0.0391(9)\\
$(\ggox {4}{5}{4}      )$ & 0.031(1) & 0.0263(5)& 0.0235(5)& 0.0195(5)& 0.059(2) & 0.050(1) & 0.045(1) & 0.0376(7)\\
$(\ggoxx{4}{5}{4}   {5})$ & 0.068(3) & 0.057(1) & 0.0509(6)& 0.043(1) & 0.068(3) & 0.057(1) & 0.0508(6)& 0.043(1) \\
$(\ggoxx{4}{5}{\ell}{m})$ & 0.057(2) & 0.048(1) & 0.0425(6)& 0.035(1) & 0.064(3) & 0.054(1) & 0.0479(7)& 0.040(1) \\
$(\ggox {4}{5}{k}      )$ & 0.047(2) & 0.0390(9)& 0.0346(5)& 0.029(1) & 0.062(3) & 0.052(1) & 0.0461(7)& 0.038(2) \\
$(\ggoI {4}{5}         )$ & 0.039(2) & 0.0322(7)& 0.0285(5)& 0.024(1) & 0.062(3) & 0.051(1) & 0.0452(7)& 0.037(2) \\
\hline
$(\gox  {5}   {5}      )$ & 0.055(1) & 0.0485(4)& 0.0425(6)& 0.036(1) & \ \ $\longleftarrow$
                                                                      & \ \ $\longleftarrow$
                                                                      & \ \ $\longleftarrow$
                                                                      & \ \ $\longleftarrow$                      \\
\end{tabular}
\end{table}

\begin{table}[p]
\squeezetable
\caption{Perturbatively renormalized pion decay constants
         $Z_A^{{\rm (P)}F} f_\pi^F$ in lattice unit.}
\label{tab:fpi1P}

\begin{tabular}{lllllllll}
\multicolumn{9}{c}{(a) $\beta = 6.0$}                                                                            \\
\hline
                          & \multicolumn{4}{c}{Gauge invariant}
                          & \multicolumn{4}{c}{Non-invariant}                                                    \\
\multicolumn{1}{c}{Operator}
                          & \multicolumn{1}{c}{$m_q a=0.030$}
                          & \multicolumn{1}{c}{$m_q a=0.020$}
                          & \multicolumn{1}{c}{$m_q a=0.010$}
                          & \multicolumn{1}{c}{$m_q a\rightarrow 0$}
                          & \multicolumn{1}{c}{$m_q a=0.030$}
                          & \multicolumn{1}{c}{$m_q a=0.020$}
                          & \multicolumn{1}{c}{$m_q a=0.010$}
                          & \multicolumn{1}{c}{$m_q a\rightarrow 0$}                                             \\
\hline
$(\ggox {4}{5}{5}      )$ & 0.0770(5)& 0.0686(6)& 0.0586(4)& 0.0495(6)& 0.0766(5)& 0.0688(6)& 0.0600(5)& 0.0519(7)\\
$(\ggoxx{4}{5}{k}   {5})$ & 0.0693(9)& 0.0630(7)& 0.0551(5)& 0.0480(6)& 0.0695(9)& 0.0635(7)& 0.0563(5)& 0.0497(6)\\
$(\ggoxx{4}{5}{k}   {4})$ & 0.066(1) & 0.0605(9)& 0.0534(5)& 0.0471(7)& 0.066(1) & 0.0609(9)& 0.0548(6)& 0.0490(8)\\
$(\ggox {4}{5}{4}      )$ & 0.058(1) & 0.053(1) & 0.0461(6)& 0.0399(9)& 0.065(1) & 0.059(1) & 0.0525(6)& 0.0463(8)\\
$(\ggoxx{4}{5}{4}   {5})$ & 0.074(2) & 0.067(1) & 0.0580(8)& 0.0502(8)& 0.073(1) & 0.067(1) & 0.0580(8)& 0.0503(9)\\
$(\ggoxx{4}{5}{\ell}{m})$ & 0.071(1) & 0.065(1) & 0.0560(7)& 0.0487(8)& 0.071(1) & 0.065(1) & 0.0566(7)& 0.0493(8)\\
$(\ggox {4}{5}{k}      )$ & 0.070(2) & 0.064(1) & 0.0546(7)& 0.0469(8)& 0.070(1) & 0.064(1) & 0.0552(7)& 0.0478(8)\\
$(\ggoI {4}{5}         )$ & 0.068(2) & 0.062(1) & 0.052(2) & 0.045(2) & 0.068(2) & 0.062(1) & 0.053(2) & 0.047(2) \\
\end{tabular}

\begin{tabular}{lllllllll}
\multicolumn{9}{c}{(b) $\beta = 6.2$}                                                                             \\
\hline
                          & \multicolumn{4}{c}{Gauge invariant}
                          & \multicolumn{4}{c}{Non-invariant}                                                     \\
\multicolumn{1}{c}{Operator}
                          & \multicolumn{1}{c}{$m_q a=0.023$}
                          & \multicolumn{1}{c}{$m_q a=0.015$}
                          & \multicolumn{1}{c}{$m_q a=0.008$}
                          & \multicolumn{1}{c}{$m_q a\rightarrow 0$}
                          & \multicolumn{1}{c}{$m_q a=0.023$}
                          & \multicolumn{1}{c}{$m_q a=0.015$}
                          & \multicolumn{1}{c}{$m_q a=0.008$}
                          & \multicolumn{1}{c}{$m_q a\rightarrow 0$}                                              \\
\hline
$(\ggox {4}{5}{5}      )$ & 0.0524(8)& 0.0454(4)& 0.0404(4)& 0.0341(6)& 0.0530(6)& 0.0464(4)& 0.0417(5)& 0.0355(9)\\
$(\ggoxx{4}{5}{k}   {5})$ & 0.049(1) & 0.0432(7)& 0.0390(6)& 0.0334(7)& 0.051(1) & 0.0442(6)& 0.0400(5)& 0.0344(9)\\
$(\ggoxx{4}{5}{k}   {4})$ & 0.049(1) & 0.0424(7)& 0.0381(5)& 0.0326(7)& 0.050(1) & 0.0435(7)& 0.0393(5)& 0.0338(8)\\
$(\ggox {4}{5}{4}      )$ & 0.045(2) & 0.0383(8)& 0.0342(8)& 0.0284(8)& 0.051(2) & 0.0433(9)& 0.0390(8)& 0.0326(6)\\
$(\ggoxx{4}{5}{4}   {5})$ & 0.055(2) & 0.046(1) & 0.0411(5)& 0.0344(9)& 0.055(2) & 0.046(1) & 0.0410(5)& 0.034(1) \\
$(\ggoxx{4}{5}{\ell}{m})$ & 0.054(2) & 0.045(1) & 0.0398(6)& 0.033(1) & 0.054(2) & 0.0453(9)& 0.0402(5)& 0.034(1) \\
$(\ggox {4}{5}{k}      )$ & 0.053(2) & 0.044(1) & 0.0392(6)& 0.033(2) & 0.054(2) & 0.0450(1)& 0.0399(6)& 0.033(2) \\
$(\ggoI {4}{5}         )$ & 0.052(3) & 0.043(1) & 0.0385(7)& 0.032(2) & 0.054(3) & 0.0444(9)& 0.0393(6)& 0.033(2) \\
\end{tabular}
\end{table}

\begin{table}[p]
\squeezetable
\caption{Nonerturbatively renormalized pion decay constants
         $Z_A^{{\rm (N)}F} f_\pi^F$ in lattice unit.}
\label{tab:fpi1N}

\begin{tabular}{lllllllll}
\multicolumn{9}{c}{(a) $\beta = 6.0$}                                                                             \\
\hline
                          & \multicolumn{4}{c}{Gauge invariant}
                          & \multicolumn{4}{c}{Non-invariant}                                                     \\
\multicolumn{1}{c}{Operator}
                          & \multicolumn{1}{c}{$m_q a=0.030$}
                          & \multicolumn{1}{c}{$m_q a=0.020$}
                          & \multicolumn{1}{c}{$m_q a=0.010$}
                          & \multicolumn{1}{c}{$m_q a\rightarrow 0$}
                          & \multicolumn{1}{c}{$m_q a=0.030$}
                          & \multicolumn{1}{c}{$m_q a=0.020$}
                          & \multicolumn{1}{c}{$m_q a=0.010$}
                          & \multicolumn{1}{c}{$m_q a\rightarrow 0$}                                              \\
\hline
$(\ggox {4}{5}{5}      )$ & 0.0770(5)& 0.0686(6)& 0.0586(4)& 0.0495(6)& 0.0731(5)& 0.0656(5)& 0.0572(5)& 0.0494(7)\\
$(\ggoxx{4}{5}{k}   {5})$ & 0.0728(9)& 0.0661(8)& 0.0579(5)& 0.0504(6)& 0.0693(9)& 0.0633(8)& 0.0562(5)& 0.0495(6)\\
$(\ggoxx{4}{5}{k}   {4})$ & 0.071(1) & 0.065(1) & 0.0575(6)& 0.0507(8)& 0.067(1) & 0.0616(9)& 0.0554(6)& 0.0495(8)\\
$(\ggox {4}{5}{4}      )$ & 0.071(1) & 0.065(1) & 0.0562(8)& 0.049(1) & 0.066(1) & 0.060(1) & 0.0535(6)& 0.0472(8)\\
$(\ggoxx{4}{5}{4}   {5})$ & 0.074(2) & 0.067(1) & 0.0585(8)& 0.0507(8)& 0.074(2) & 0.067(1) & 0.0585(8)& 0.0507(9)\\
$(\ggoxx{4}{5}{\ell}{m})$ & 0.075(1) & 0.068(1) & 0.0594(8)& 0.0516(9)& 0.073(1) & 0.067(1) & 0.0582(7)& 0.0507(8)\\
$(\ggox {4}{5}{k}      )$ & 0.075(2) & 0.068(1) & 0.0586(8)& 0.0505(8)& 0.072(1) & 0.066(1) & 0.0566(7)& 0.0490(8)\\
$(\ggoI {4}{5}         )$ & 0.074(2) & 0.068(2) & 0.057(2) & 0.050(2) & 0.070(2) & 0.064(1) & 0.055(2) & 0.048(2) \\
\end{tabular}

\begin{tabular}{lllllllll}
\multicolumn{9}{c}{(b) $\beta = 6.2$}                                                                             \\
\hline
                          & \multicolumn{4}{c}{Gauge invariant}
                          & \multicolumn{4}{c}{Non-invariant}                                                     \\
\multicolumn{1}{c}{Operator}
                          & \multicolumn{1}{c}{$m_q a=0.023$}
                          & \multicolumn{1}{c}{$m_q a=0.015$}
                          & \multicolumn{1}{c}{$m_q a=0.008$}
                          & \multicolumn{1}{c}{$m_q a\rightarrow 0$}
                          & \multicolumn{1}{c}{$m_q a=0.023$}
                          & \multicolumn{1}{c}{$m_q a=0.015$}
                          & \multicolumn{1}{c}{$m_q a=0.008$}
                          & \multicolumn{1}{c}{$m_q a\rightarrow 0$}                                              \\
\hline
$(\ggox {4}{5}{5}      )$ & 0.0524(8)& 0.0454(4)& 0.0404(4)& 0.0341(6)& 0.0513(6)& 0.0450(4)& 0.0404(5)& 0.0344(9)\\
$(\ggoxx{4}{5}{k}   {5})$ & 0.051(1) & 0.0449(8)& 0.0405(6)& 0.0347(7)& 0.051(1) & 0.0441(6)& 0.0400(5)& 0.0344(9)\\
$(\ggoxx{4}{5}{k}   {4})$ & 0.052(1) & 0.0448(8)& 0.0403(5)& 0.0345(7)& 0.050(1) & 0.0439(7)& 0.0397(5)& 0.0342(8)\\
$(\ggox {4}{5}{4}      )$ & 0.053(2) & 0.0451(9)& 0.0403(9)& 0.0335(9)& 0.052(2) & 0.0440(9)& 0.0396(8)& 0.0331(6)\\
$(\ggoxx{4}{5}{4}   {5})$ & 0.055(2) & 0.047(1) & 0.0414(5)& 0.035(1) & 0.055(2) & 0.046(1) & 0.0413(5)& 0.035(1) \\
$(\ggoxx{4}{5}{\ell}{m})$ & 0.056(2) & 0.047(1) & 0.0418(6)& 0.035(1) & 0.055(2) & 0.046(1) & 0.0412(6)& 0.034(1) \\
$(\ggox {4}{5}{k}      )$ & 0.056(3) & 0.047(1) & 0.0415(7)& 0.034(2) & 0.055(2) & 0.046(1) & 0.0407(6)& 0.034(2) \\
$(\ggoI {4}{5}         )$ & 0.056(3) & 0.047(1) & 0.0412(7)& 0.034(2) & 0.055(3) & 0.045(1) & 0.0401(7)& 0.033(2) \\
\end{tabular}
\end{table}

\begin{table}[p]
\caption{Pion mass squared $(m_\pi^F)^2$ in GeV$^2$.}
\label{tab:mass5}

\mediumtext

\begin{tabular}{lllllll}
                      & \multicolumn{3}{c}{Gauge invariant}
                      & \multicolumn{3}{c}{Non-invariant}                               \\
\multicolumn{1}{c}{Operator}
                      & \multicolumn{1}{c}{$\beta=6.0$}
                      & \multicolumn{1}{c}{$\beta=6.2$}
                      & \multicolumn{1}{c}{$a\rightarrow 0$}
                      & \multicolumn{1}{c}{$\beta=6.0$}
                      & \multicolumn{1}{c}{$\beta=6.2$}
                      & \multicolumn{1}{c}{$a\rightarrow 0$}                            \\
\hline
$(\gox {5}{5}      )$ & 0.0066(9)& 0.003(3) & 0.000(6) & \ $\longleftarrow$
                                                       & \ $\longleftarrow$
                                                       & \ $\longleftarrow$             \\
$(\goxx{5}{k}   {5})$ & 0.085(2) & 0.042(4) & 0.002(8) & 0.085(2) & 0.042(4) & 0.002(8) \\
$(\goxx{5}{k}   {4})$ & 0.118(3) & 0.061(4) & 0.006(9) & 0.118(3) & 0.061(4) & 0.005(9) \\
$(\gox {5}{4}      )$ & 0.147(6) & 0.074(5) & 0.000(10)& 0.147(6) & 0.074(5) & 0.000(10)\\
$(\goxx{5}{4}   {5})$ & 0.084(2) & 0.042(3) & 0.001(6) & 0.084(2) & 0.041(3) & 0.000(6) \\
$(\goxx{5}{\ell}{m})$ & 0.118(3) & 0.059(4) & 0.002(8) & 0.118(3) & 0.059(4) & 0.002(8) \\
$(\gox {5}{k}      )$ & 0.146(4) & 0.074(4) & 0.004(9) & 0.146(4) & 0.073(5) & 0.000(10)\\
$(\goI {5}         )$ & 0.178(5) & 0.092(4) & 0.009(9) & 0.178(6) & 0.091(5) & 0.010(10)\\
\end{tabular}
\end{table}

\narrowtext

\begin{table}[p]
\caption{Perturbatively renormalized pion decay constants
         $Z_A^{{\rm (P)}F} f_\pi^F$ in MeV.  The bottom line
         shows results obtained with the pion operator in the $\xi_5$
         channel $f_\pi^{(P)5}$.}
\label{tab:fpi5P}

\begin{tabular}{lllllll}
                          & \multicolumn{3}{c}{Gauge invariant}
                          & \multicolumn{3}{c}{Non-invariant}             \\
\multicolumn{1}{c}{Operator}
                          & \multicolumn{1}{c}{$\beta=6.0$}
                          & \multicolumn{1}{c}{$\beta=6.2$}
                          & \multicolumn{1}{c}{$a\rightarrow 0$}
                          & \multicolumn{1}{c}{$\beta=6.0$}
                          & \multicolumn{1}{c}{$\beta=6.2$}
                          & \multicolumn{1}{c}{$a\rightarrow 0$}          \\
\hline
$(\ggox {4}{5}{5}      )$ & 95(2) & 92(3) & 89(6) &100(2) & 96(4) & 92(7) \\
$(\ggoxx{4}{5}{k}   {5})$ & 92(1) & 90(3) & 88(5) & 96(2) & 93(3) & 90(6) \\
$(\ggoxx{4}{5}{k}   {4})$ & 91(2) & 88(3) & 85(5) & 94(2) & 91(2) & 88(5) \\
$(\ggox {4}{5}{4}      )$ & 77(2) & 77(3) & 77(5) & 89(2) & 88(2) & 87(4) \\
$(\ggoxx{4}{5}{4}   {5})$ & 97(2) & 93(3) & 89(7) & 97(2) & 93(4) & 88(7) \\
$(\ggoxx{4}{5}{\ell}{m})$ & 94(2) & 90(3) & 86(7) & 95(2) & 91(4) & 86(7) \\
$(\ggox {4}{5}{k}      )$ & 90(2) & 88(4) & 85(9) & 92(2) & 89(5) & 87(9) \\
$(\ggoI {4}{5}         )$ & 87(4) & 86(5) & 85(10)& 90(4) & 88(5) & 86(11)\\
\hline
$(\gox  {5}   {5}      )$ & 98(1) & 94(3) & 89(6) & $\longleftarrow$
                                                  & $\longleftarrow$
                                                  & $\longleftarrow$      \\
\end{tabular}
\end{table}

\begin{table}[p]
\caption{Non-perturbatively renormalized pion decay constants
         $Z_A^{{\rm (N)}F} f_\pi^F$ in MeV unit.  The bottom
         line for $f_\pi^{(P) 5}$ is reproduced from
         Table~\ref{tab:fpi5P} for convenience.}
\label{tab:fpi5N}

\begin{tabular}{lllllll}
                          & \multicolumn{3}{c}{Gauge invariant}
                          & \multicolumn{3}{c}{Non-invariant}             \\
\multicolumn{1}{c}{Operator}
                          & \multicolumn{1}{c}{$\beta=6.0$}
                          & \multicolumn{1}{c}{$\beta=6.2$}
                          & \multicolumn{1}{c}{$a\rightarrow 0$}
                          & \multicolumn{1}{c}{$\beta=6.0$}
                          & \multicolumn{1}{c}{$\beta=6.2$}
                          & \multicolumn{1}{c}{$a\rightarrow 0$}          \\
\hline
$(\ggox {4}{5}{5}      )$ & 95(2) & 92(3) & 89(6) & 95(2) & 93(3) & 91(7) \\
$(\ggoxx{4}{5}{k}   {5})$ & 97(2) & 94(3) & 90(6) & 95(2) & 93(3) & 90(6) \\
$(\ggoxx{4}{5}{k}   {4})$ & 98(2) & 93(3) & 89(6) & 95(2) & 92(2) & 89(5) \\
$(\ggox {4}{5}{4}      )$ & 94(2) & 90(3) & 87(6) & 91(2) & 89(2) & 88(4) \\
$(\ggoxx{4}{5}{4}   {5})$ & 98(2) & 94(3) & 90(7) & 98(2) & 93(4) & 89(7) \\
$(\ggoxx{4}{5}{\ell}{m})$ & 99(2) & 94(3) & 89(7) & 98(2) & 93(4) & 88(8) \\
$(\ggox {4}{5}{k}      )$ & 97(2) & 93(5) & 89(9) & 94(2) & 91(5) & 88(9) \\
$(\ggoI {4}{5}         )$ & 96(5) & 92(5) & 89(11)& 92(4) & 90(5) & 88(11)\\
\hline
$(\gox  {5}   {5}      )$ & 98(1) & 94(3) & 89(6) & $\longleftarrow$
                                                  & $\longleftarrow$
                                                  & $\longleftarrow$      \\
\end{tabular}
\end{table}

\end{document}